\documentclass[epj]{svjour}
\usepackage{graphicx}
\input{epsf}
\usepackage{amssymb}

\begin{document}

\title{Time evolution of Wikipedia network ranking}

\author{
Young-Ho Eom\inst{1}
\and
Klaus M. Frahm\inst{1}
\and
Andr\'as Bencz\'ur\inst{2}
\and
Dima L. Shepelyansky\inst{1}
}
\institute{
Laboratoire de Physique Th\'eorique du CNRS, IRSAMC, 
Universit\'e de Toulouse, UPS, 31062 Toulouse, France\\
\and
Informatics Laboratory, Institute for Computer Science and Control, \\
Hungarian Academy of Sciences (MTA SZTAKI), Pf. 63, H-1518 Budapest, Hungary
}

\titlerunning{Time evolution of Wikipedia network ranking}
\authorrunning{Y.-H.Eom, K.M.Frahm, A.Benc\'zur  and D.L.Shepelyansky}

\abstract{We study the time evolution of ranking and spectral
properties of   the Google matrix of English Wikipedia hyperlink network
during years 2003 - 2011. The statistical properties 
of ranking of Wikipedia articles via
PageRank and CheiRank probabilities,
as well as the matrix spectrum, are shown to be
stabilized  for 2007 - 2011.
A special emphasis is done on ranking of 
Wikipedia personalities
and universities. We show that PageRank selection
is dominated by politicians while 2DRank, which 
combines PageRank and CheiRank, gives more accent
on personalities of arts. 
The Wikipedia
PageRank of universities recovers
80 percents of top universities of Shanghai ranking
during the considered time period.
}


\date{Received: April 24, 2013}

\maketitle

\section{ Introduction}

At present Wikipedia \cite{wikipedia}
has become the world largest Encyclopedia
with open public access to its content. 
A recent review \cite{finn} 
represents a detailed description of publications and scientific research 
on information storage at Wikipedia and its classification.
Wikipedia contains
an enormous amount of information 
and, in a certain sense, the problem of information 
arrangement and retrieval
from its contain starts
to remind similar information problems
in the Library of Babel
described by Jorge Luis Borges \cite{borges}. 
The hyperlinks 
between Wikipedia articles 
represent a directed network which 
reminds a structure of the World Wide Web (WWW).
Hence, the mathematical tools developed
for WWW search engines, based on the Markov chains \cite{markov},
Perron-Frobenius operators \cite{mbrin}
and the PageRank algorithm 
of the corresponding Google matrix \cite{brin,meyerbook},
give solid mathematical grounds for analysis of
information flow on the Wikipedia network.
In this work we perform the Google matrix
analysis of  Wikipedia network of English articles
 extending the results presented in 
\cite{zzswiki,2dmotor},\cite{twitter,wikispectrum}.
The main new element of this work
is the study of time evolution
of Wikipedia network during the years
2003 to 2011. We analyze how the ranking
of Wikipedia articles 
and the spectrum of the Google matrix $G$ of Wikipedia 
are changed during this period. 

The directed network of Wikipedia articles
is const\-ructed in a standard way: a directed link 
is formed from an article $j$ to an article $i$
when $j$ quotes $i$ and an element $A_{ij}$ of
the adjacency matrix is taken to be unity when
there is such a link and zero in absence of link.
The columns with only zero elements ({\em dangling nodes})
are replaced by columns with $1/N$ with $N$ being the matrix size.
The elements of other columns are renormalized
in such a way that their sum becomes equal to unity
($\sum_j S_{ij}=1$, $S_{ij}=A_{ij}/\sum_i A_{ij}$).
Thus we obtain the matrix $S_{ij}$ of Markov transitions.
Then the Google matrix of the network takes the form
\cite{brin,meyerbook}:
\begin{equation}
   G_{ij} = \alpha  S_{ij} + (1-\alpha)/N \;\; .
\label{eq1} 
\end{equation} 
The damping parameter $\alpha$ in the WWW context 
describes the probability 
$(1-\alpha)$ to jump to any node for a random surfer. 
For WWW the Google search engine uses 
$\alpha \approx 0.85$ \cite{meyerbook}.
The matrix $G$ belongs to the class of Perron-Frobenius 
operators \cite{mbrin,meyerbook},
its largest eigenvalue 
is $\lambda = 1$ and other eigenvalues have $|\lambda| \le \alpha$. 
The right eigenvector at $\lambda = 1$, which is called the PageRank, 
has real nonnegative elements $P(i)$
and gives a probability $P(i)$ to find a random surfer at site $i$. 
It is possible to rank all nodes
in a decreasing order of PageRank probability $P(K(i))$
so that the PageRank index $K(i)$ counts all $N$ nodes
$i$ according their ranking, placing 
the most popular articles or nodes
at the top values $K=1,2,3 ...$.

Due to the gap $1-\alpha\approx 0.15$ between 
the largest eigenvalue $\lambda=1$ and 
other eigenvalues the PageRank algorithm permits an efficient 
and simple determination of the 
PageRank by the power iteration method \cite{meyerbook}. 
It is also possible to use the powerful Arnoldi method 
\cite{arnoldibook,golub},\cite{ulamfrahm}
to compute efficiently the eigenspectrum $\lambda_i$ 
of the Google matrix:
\begin{equation}
   \sum_{k=1}^N G_{jk} \psi_i(k) = \lambda_i \psi_i(j) \;\; .
\label{eq2} 
\end{equation} 
The Arnoldi method allows to find a several thousands
of eigenvalues $\lambda_i$ with maximal 
$|\lambda|$ for a matrix size
$N $ as large as a few tens of millions \cite{twitter,wikispectrum},
\cite{ulamfrahm,univuk}. 
Usually, at $\alpha=1$ the largest eigenvalue $\lambda=1$ is 
highly degenerate \cite{univuk} due to many invariant subspaces which 
define many independent Perron-Frobenius operators 
providing (at least)
one eigenvalue $\lambda=1$.

In addition to a given directed network $A_{ij}$
it is useful to analyze an inverse network
with inverted direction of links with
elements of adjacency matrix $A_{ij} \rightarrow A_{ji}$.
The Google matrix $G^*$ of the inverse network 
is then constructed via corresponding matrix $S^*$
according to the relations
(\ref{eq1}) using the same value of $\alpha$ as for the $G$ matrix.
This time inversion approach was used in
\cite{fogaras,hristidis} but the statistical
properties and correlations between direct and inversed
ranking were not analyzed there.
In \cite{alik}, on an example of the Linux Kernel network,
it was shown that this approach allows to obtain 
an additional interesting characterization
of information flow on directed networks.
Indeed, the right eigenvector of $G^*$ at eigenvalue $\lambda=1$
gives a probability $P^*(i)$,
called CheiRank vector \cite{zzswiki}. 
It determines
a complementary rank index $K^*(i)$ of network nodes
in a decreasing order of probability $P^*(K^*(i))$
\cite{zzswiki,2dmotor},\cite{twitter,alik}.
It is known that the PageRank probability 
is proportional to the number of ingoing links
characterizing how popular or known is a given node.
In a similar way the 
CheiRank probability is proportional to the 
number of outgoing links highlighting the node communicativity
(see e.g. \cite{meyerbook,donato},
\cite{upfal,litvak},\cite{zzswiki,2dmotor}).
The statistical properties of distribution of indexes 
$K(i),K^*(i)$
on the PageRank-CheiRank plane are described 
in \cite{2dmotor}.

In this work we apply the above mathematical methods
to the analysis of time evolution of Wikipedia network ranking
using English Wikipedia snapshots dated
by December 31 of years 2003, 2005, 2007, 2009, 2011.
In addition we use the snapshot of August 2009 (200908)
analyzed in \cite{zzswiki}. The parameters of networks
with the number of articles (nodes) $N$, number of links
$N_\ell$ and other information are given in Tables 1,2 
with the description of notations given in Appendix.

The paper is composed as following:
the statistical properties of PageRank and CheiRank
are analyzed in Section 2, ranking of Wikipedia personalities 
and universities 
are considered in Sections 3, 4 respectively,
the properties of spectrum of Google matrix are
considered in Section 5, the discussion of the results
is presented in Section 6,  Appendix
Section 7 gives network parameters.

\section{CheiRank versus PageRank}

\begin{figure}
\begin{center}
\includegraphics[clip=true,width=3.6cm,angle=-90]{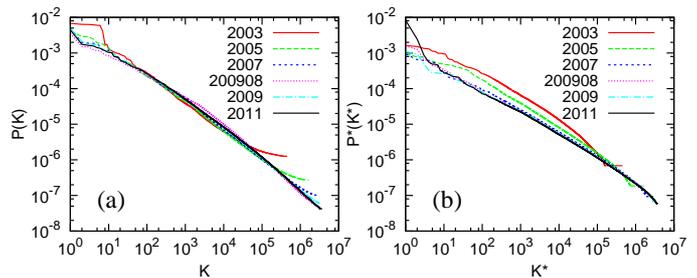}
\vglue -0.1cm
\caption{PageRank probability $P(K)$ (left panel)
and  CheiRank probability $P^*(K^*)$ (right panel)
are shown as a function of the corresponding rank
indexes $K$ and $K^*$ for English Wikipedia 
articles at years 2003, 2005, 2007,  200908, 2009, 2011;
here the damping factor is $\alpha = 0.85$.
}
\label{fig1}
\end{center}
\end{figure}

The dependencies of PageRank and CheiRank probabilities $P(K)$
and $P^*(K^*)$ on their indexes $K$, $K^*$
at different years are shown in Fig.~\ref{fig1}.
The top positions of $K$ are occupied by countries
starting from {\it United States} while at the top 
positions of $K^*$ we find various lists 
(e.g.  geographical names, prime ministers etc.; 
in 2011 we have appearance of lists  of lists). 
Indeed, the countries accumulate links
from all types of human activities and nature,
that make them the most popular Wikipedia articles,
while lists have the largest number
of outgoing links making them the most communicative
articles. 

The data of Fig.~\ref{fig1} show that 
the global behavior of $P(K)$
remains stable from 2007 to 2011.
Indeed, the decay of probability curves 
$P(K)$ is very similar and 4 curves
are practically overlapped in $K <10^6$. Also 
the probability decay $P^*(K^*)$ is described by
curves been very close to each other for
the time interval 2007 - 2009
while at 2011 we see the appearance of
peak at $1\leq K^* < 10$. 
This peak is related to introduction of 
lists of lists which were absent at earlier years.
At the same time the behavior of $P^*(K^*)$ in the range
$10 \leq K^* \leq 10^6$ remains  stable  
for 2007 - 2011. Indeed, we see that the probability curves
are very close to each other as it is well visible in Fig.~1.
However, for a quantitative analysis
one needs to consider overlap of articles 
in the top ranking at different years.
We discuss this point below.

\begin{figure}
\begin{center}
\includegraphics[clip=true,width=3.7cm,angle=-90]{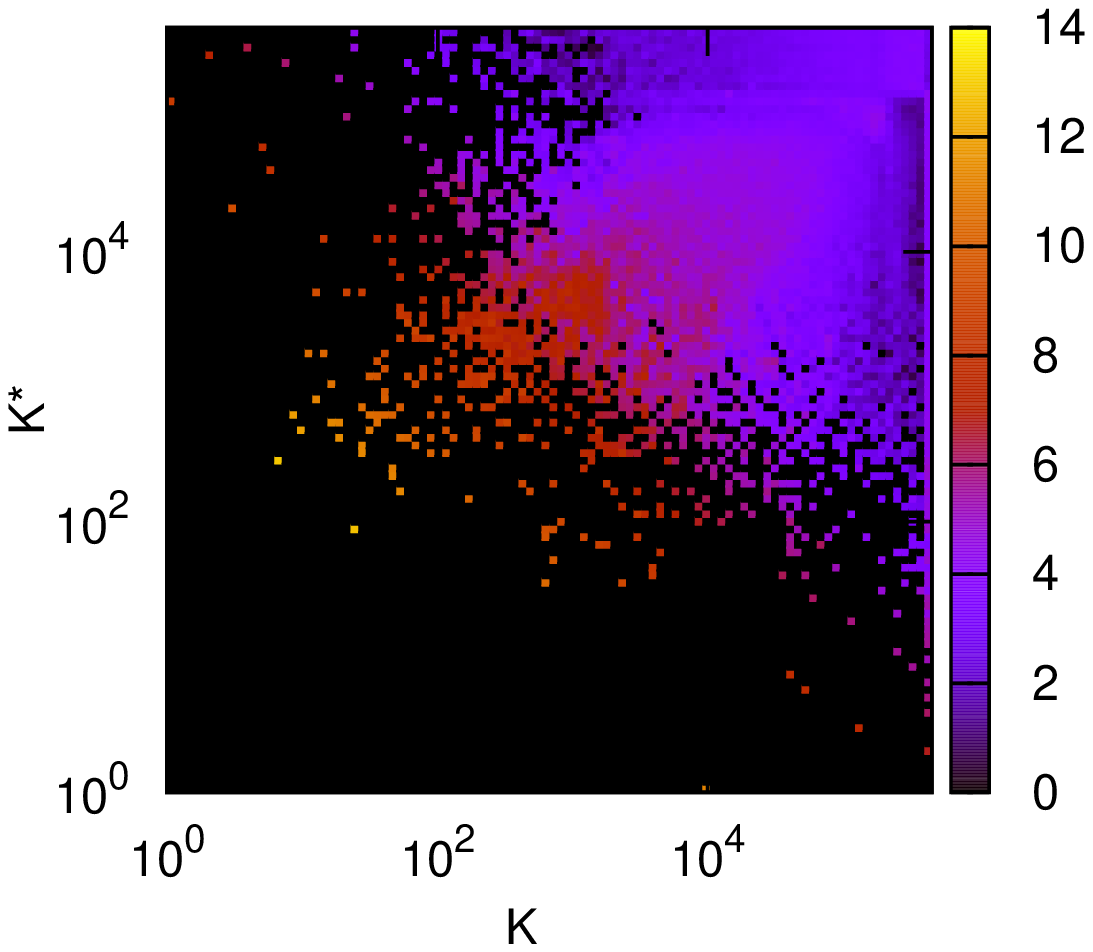}
\includegraphics[clip=true,width=3.7cm,angle=-90]{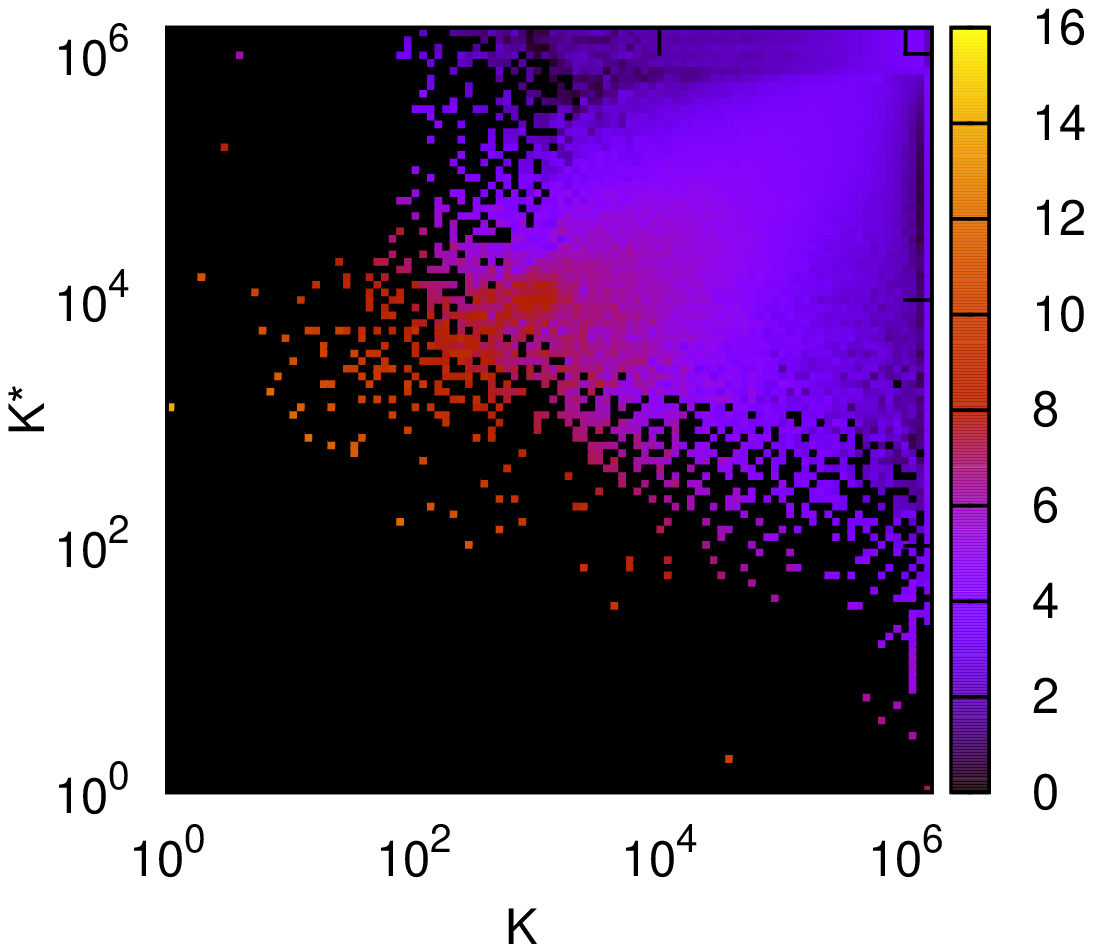}\\
\includegraphics[clip=true,width=3.7cm,angle=-90]{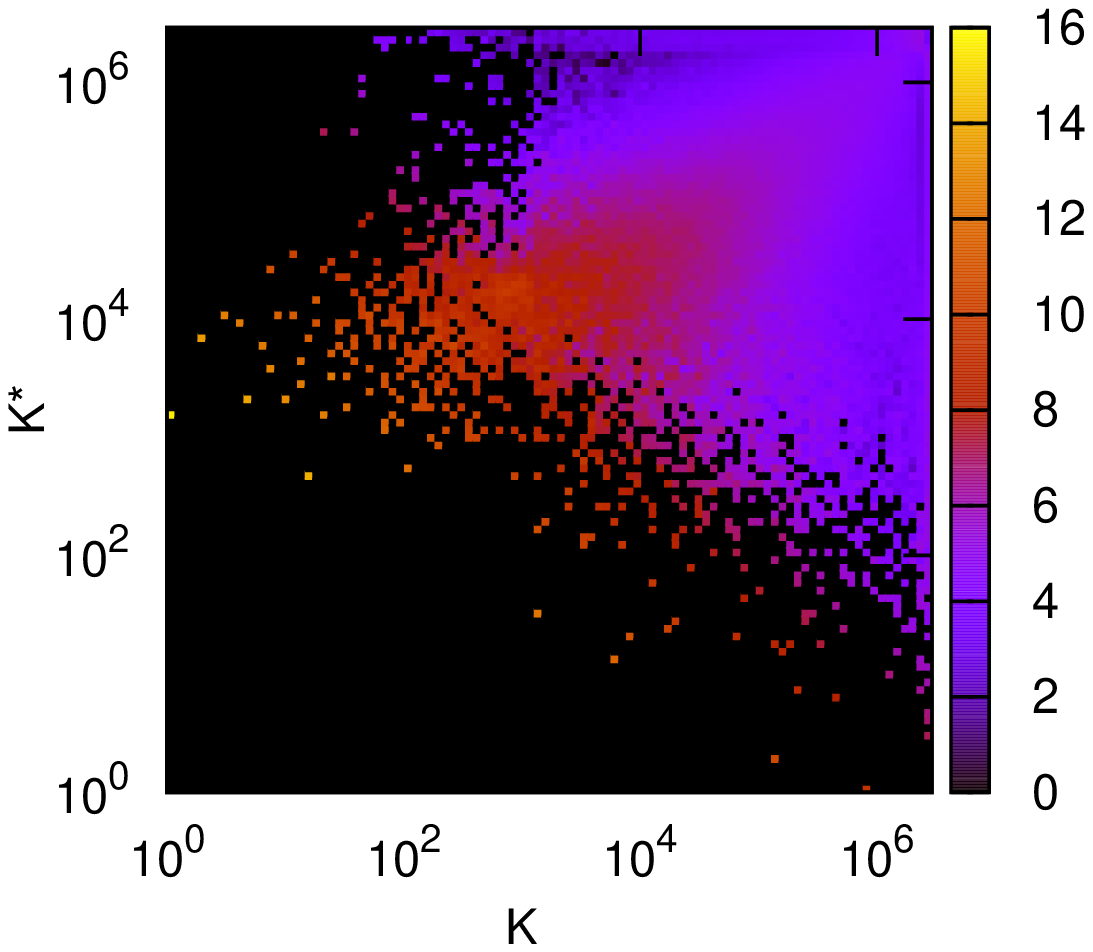}
\includegraphics[clip=true,width=3.7cm,angle=-90]{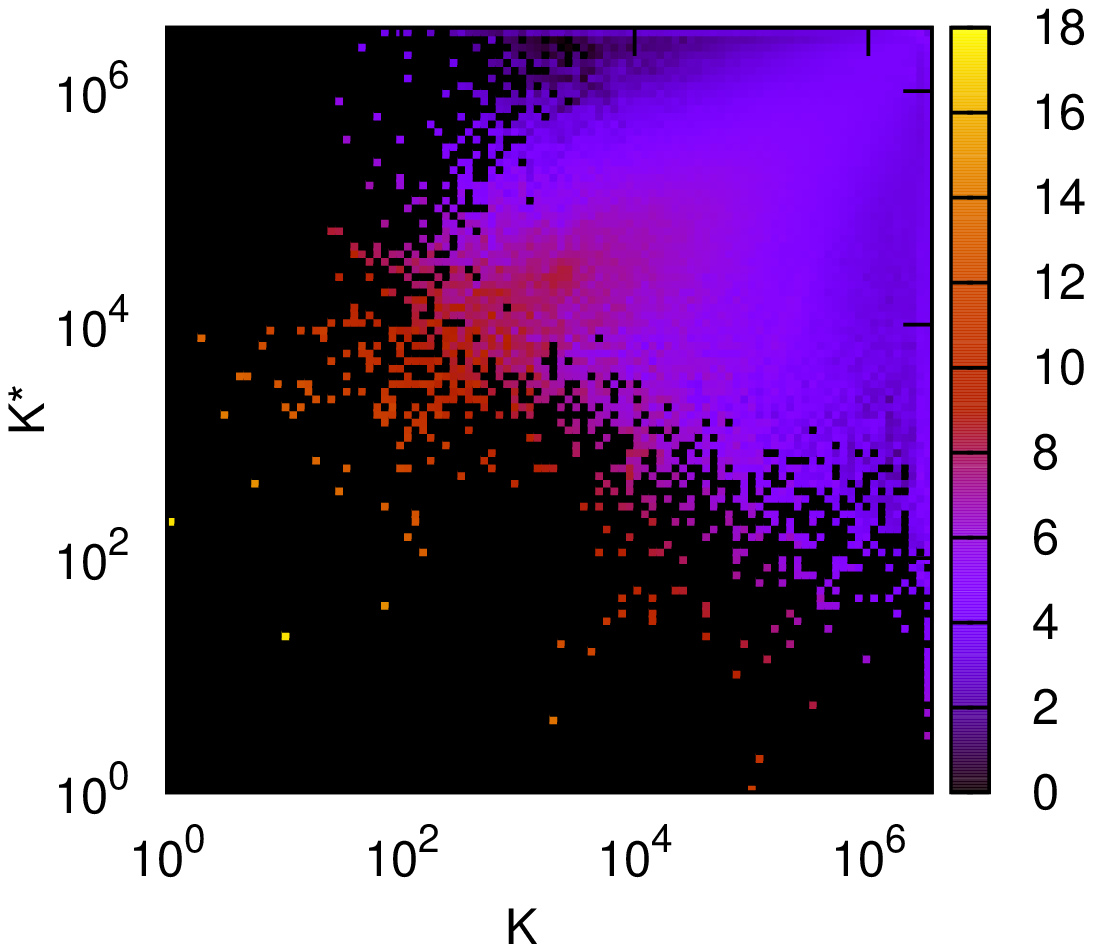}\\
\includegraphics[clip=true,width=3.7cm,angle=-90]{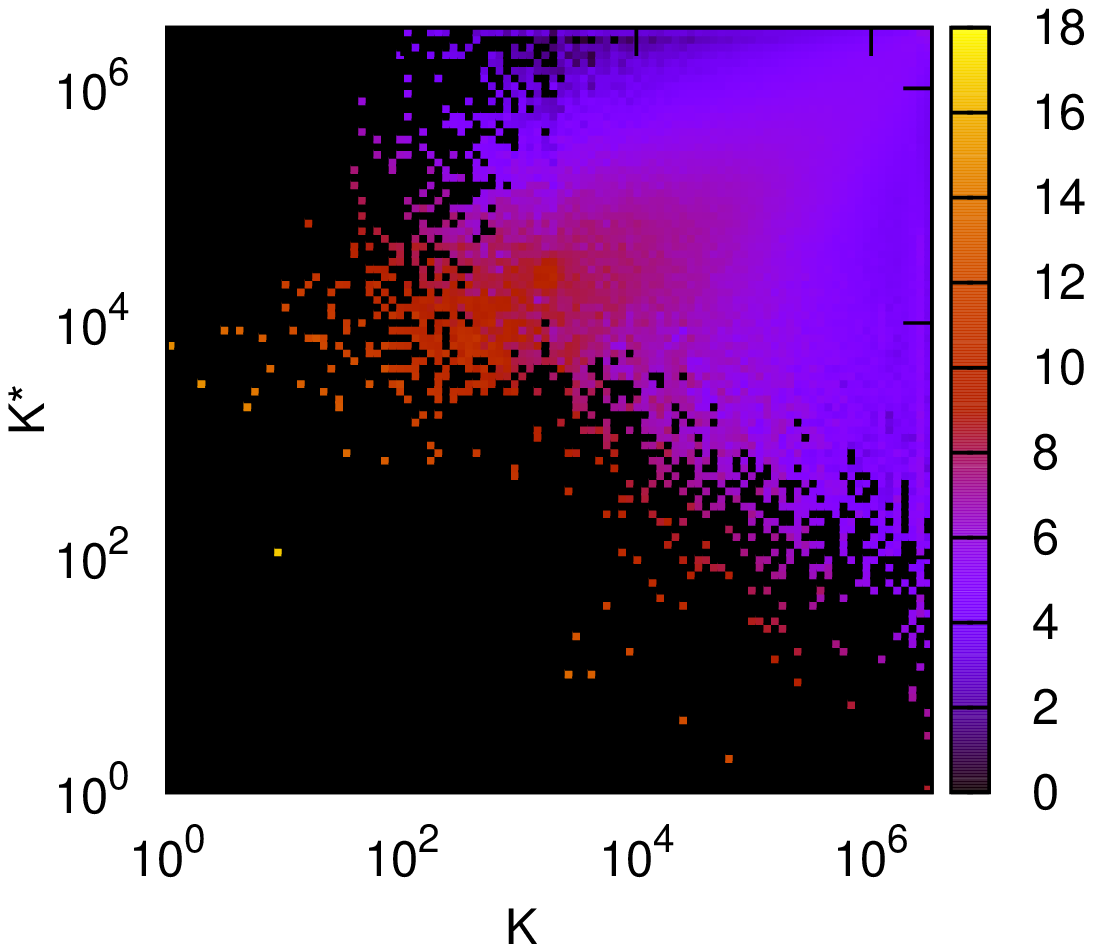}
\includegraphics[clip=true,width=3.7cm,angle=-90]{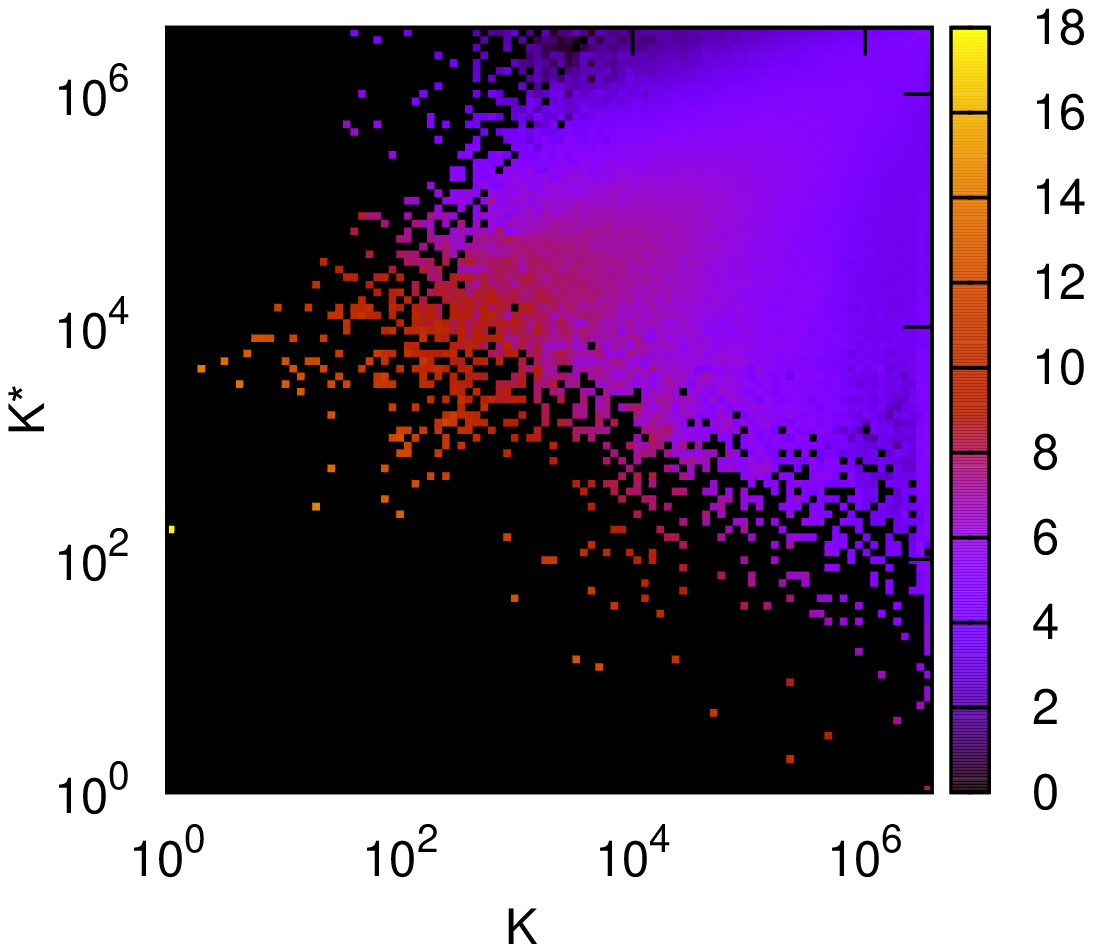}
\vglue -0.1cm
\caption{Density of Wikipedia articles 
in the CheiRank versus PageRank plane at different years.
Color is proportional to  logarithm of density 
changing from minimal nonzero density (dark) to maximal one
(white), zero density is shown by black
(distribution is computed for $100 \times 100$
cells equidistant in logarithmic scale; bar shows color 
variation of natural logarithm of density);
left column panels are for years 2003, 2007, 200908
and right column panels are for 2005, 2009, 2011
(from top to bottom). 
}
\label{fig2}
\end{center}
\end{figure}

Each article $i$ has its PageRank and CheiRank indexes
$K(i)$, $K^*(i)$ so that all articles are distributed on 
two-dimensional plane of PageRank-CheiRank indexes.
Following \cite{zzswiki,2dmotor}, we present the density
of articles in the 2D plane $(K,K^*)$ in Fig.~\ref{fig2}.
The density is computed for $100 \times 100$
logarithmically equidistant cells which cover the whole
plane  $(K,K^*)$ for each year. Qualitatively we find that the 
density distribution is globally stable for
years 2007-2011 even if definitely there are articles 
which change their location in 2D plane. For example,
we see an appearance of a mountain like ridge
of probability located approximately along a line
$\ln K^* \approx \ln K + 4.6$ that indicates
the presence of correlation between
$P(K(i))$ and $P^*(K^*(i))$. 
Also the form of density distributions
looks to be similar at all years studied
even if individual articles change their positions.

Following \cite{zzswiki,2dmotor,alik},
we characterize the interdependence of
PageRank and CheiRank vectors by the correlator 
\begin{equation}
  \kappa =N \sum^N_{i=1} P(K(i)) P^*(K^*(i)) - 1 \;\; .
\label{eq3} 
\end{equation}

\begin{figure}
\begin{center}
\includegraphics[clip=true,width=5.5cm,angle=-90]{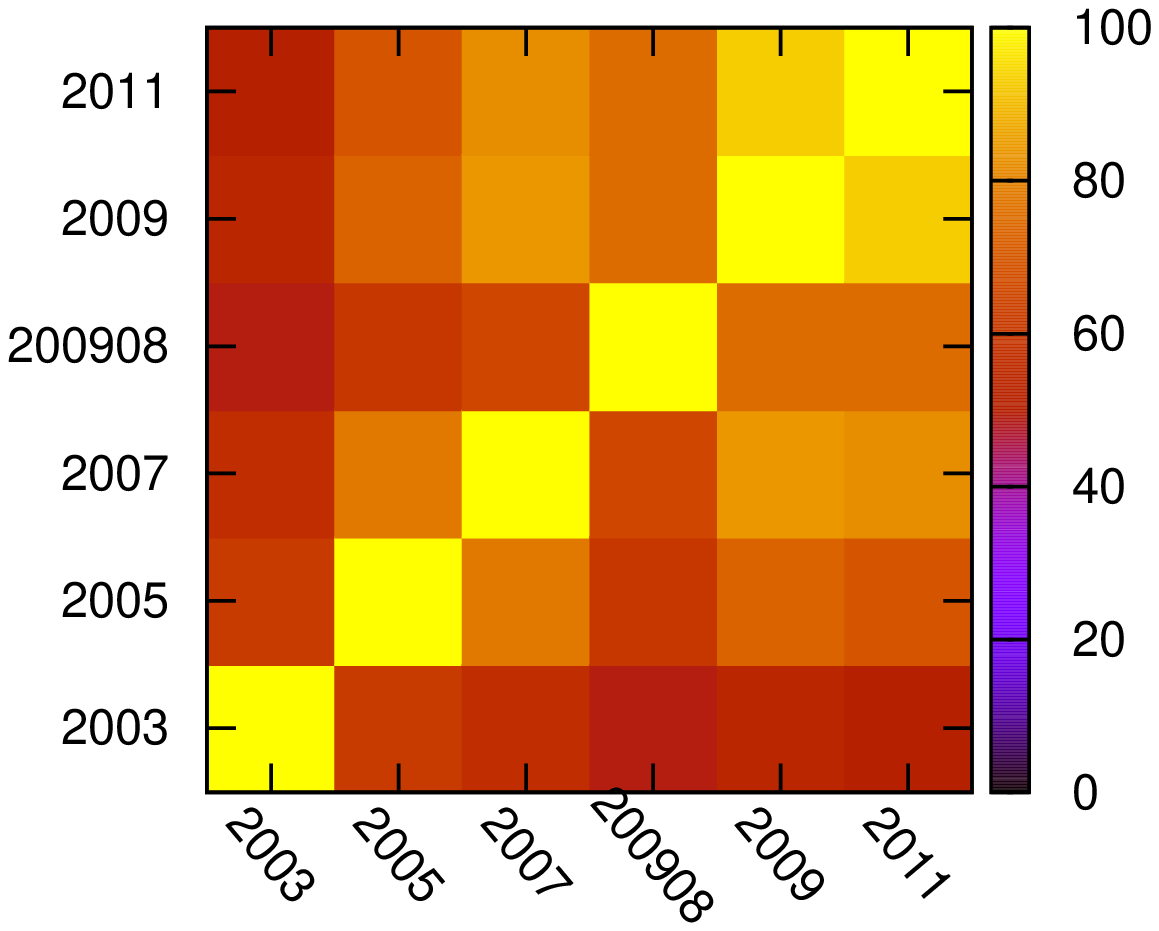}\\
\includegraphics[clip=true,width=5.5cm,angle=-90]{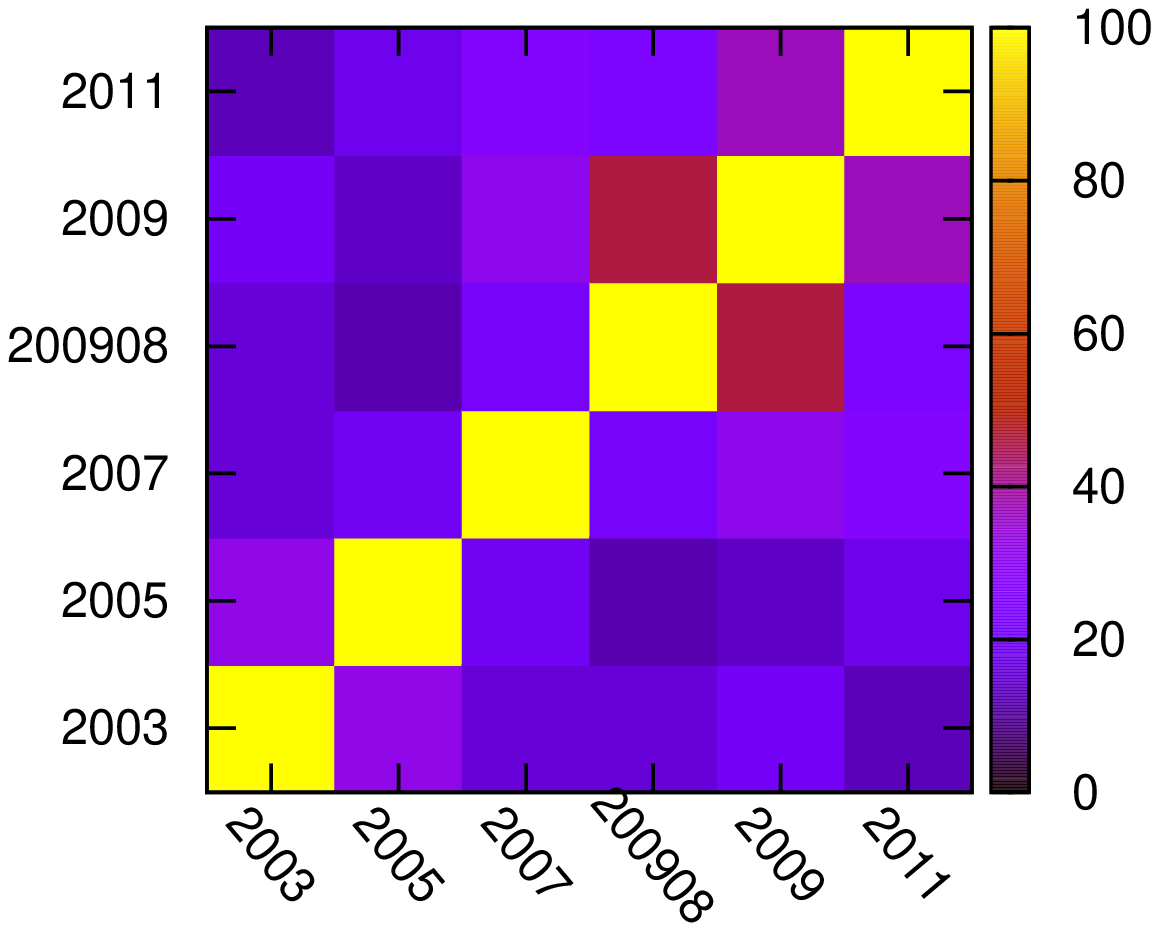}\\
\includegraphics[clip=true,width=5.5cm,angle=-90]{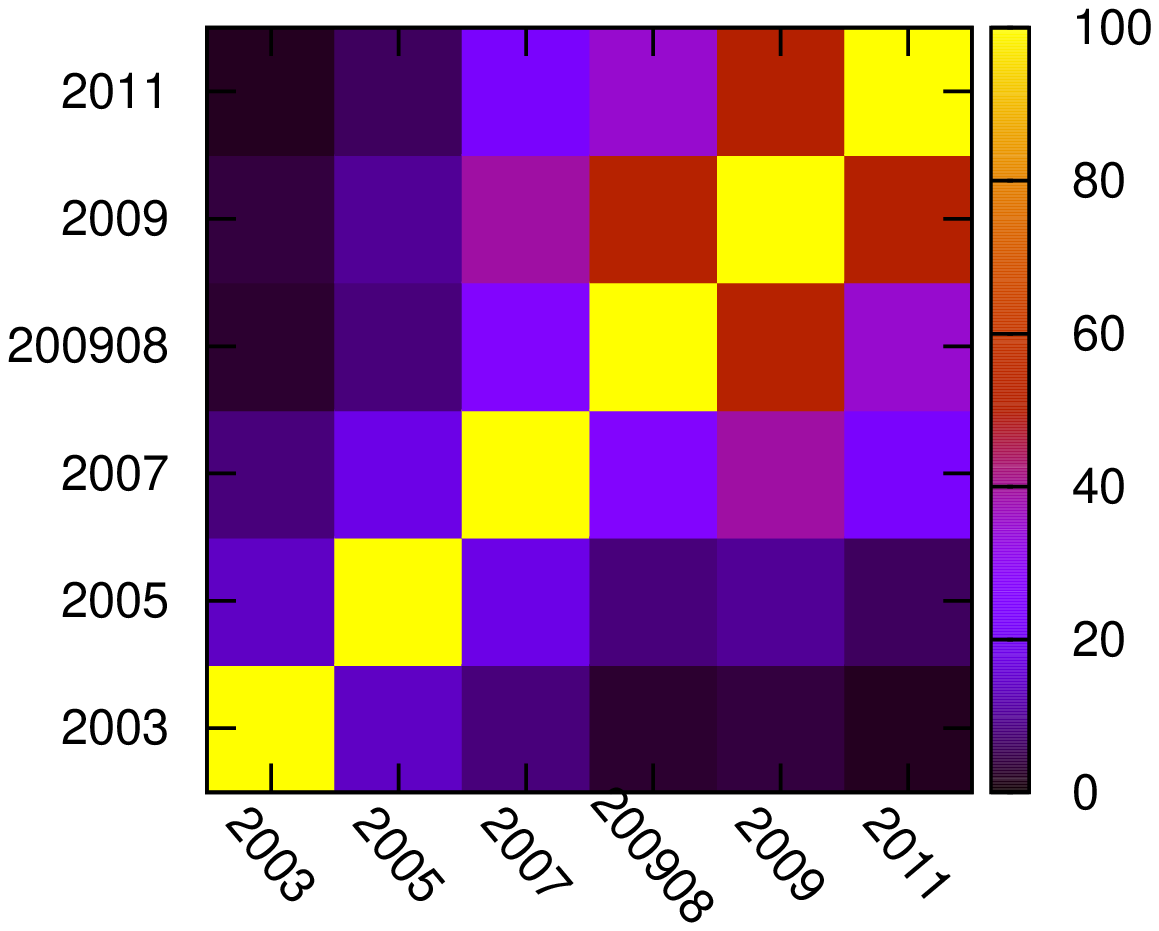}
\vglue -0.1cm
\vglue -0.1cm
\caption{ Number of the same overlapped articles 
between top 100 Wikipedia articles
at different years for ranking by PageRank (top panel),
2DRank (middle panel) and CheiRank (bottom panel).
}
\label{fig3}
\end{center}
\end{figure}

We find the following values of
the correlator at various time slots:
$\kappa=2.837 (2003)$, $3.894 (2005)$, $4.121 (2007)$,
$4.084 (200908)$, $6.629 (2009)$, $5.391 (2011)$.
During that period the size of the network 
increased almost by $10$ times while
$\kappa$ increased less than $2$ times.
The root mean square variation around the average value $\kappa=4.49$
is relatively modest being $\delta \kappa =1.2$. 
The stability of $\kappa$ is 
especially visible in comparison with other networks.
Indeed, for the network of University of Cambridge
we have $\kappa =1.71$ in 2006 and $30$ in 2011 \cite{2dmotor}.
This confirms the stability of the correlator $\kappa$
during the time evolution of the Wikipedia network.

To analyze the stability of ranking in a more quantitative manner,
we determine the number of the same overlapping articles 
at different years among the top 100 articles in
PageRank, 2DRank and CheiRank. The dependence of this overlap
characteristic $N_{ovlap}$ on different years 
is shown in Fig.~\ref{fig3}. For PageRank we have 
the lowest value $N_{ovlap} \approx 40$ and for
the time period $2007-2011$ we have this value mainly
in the range $60-80$ confirming the stability of top 100
articles of PageRank. For 2DRank we have smaller values of
$N_{ovlap}$ which are located mainly in the range
$30 - 50$ for the period $2007-2011$
with overlap drop to $10$ between 2003 and 2011.
For CheiRank we find approximately the same
of overlap as for 2DRank for years 2007 - 2011.
However, e.g. for years 2003 vs 2011 the overlap 
for CheiRank drops significantly down to the minimal value
$N_{ovlap}=2$.
We attribute this to significant fluctuations of top
100 CheiRank probabilities  especially
visible in Fig.~\ref{fig1} for years 2003, 2005.
The significant values of overlap
parameter $N_{ovlap}$ for years 2007-2011
indicate the stabilization of rank distributions
in this period.

In the next two Sections we analyze the 
time variation of ranking of personalities and
universities.

\section{Ranking of personalities}

To analyze the time evolution of 
ranking of Wikipedia personalities 
(persons or humans) we chose the top 100
persons appearing in the ranking list
of Wikipedia 200908 given in \cite{zzswiki}
in order of PageRank, CheiRank and 2DRank.
We remind that 2DRank $K_2$ is obtained by
counting nodes in order of their appearance
on ribs of squares in $(K,K^*)$ plane
with their size growing from $K=1$ to $K=N$ \cite{zzswiki}.

\begin{figure}
\begin{center}
\includegraphics[clip=true,width=3.9cm,angle=-90]{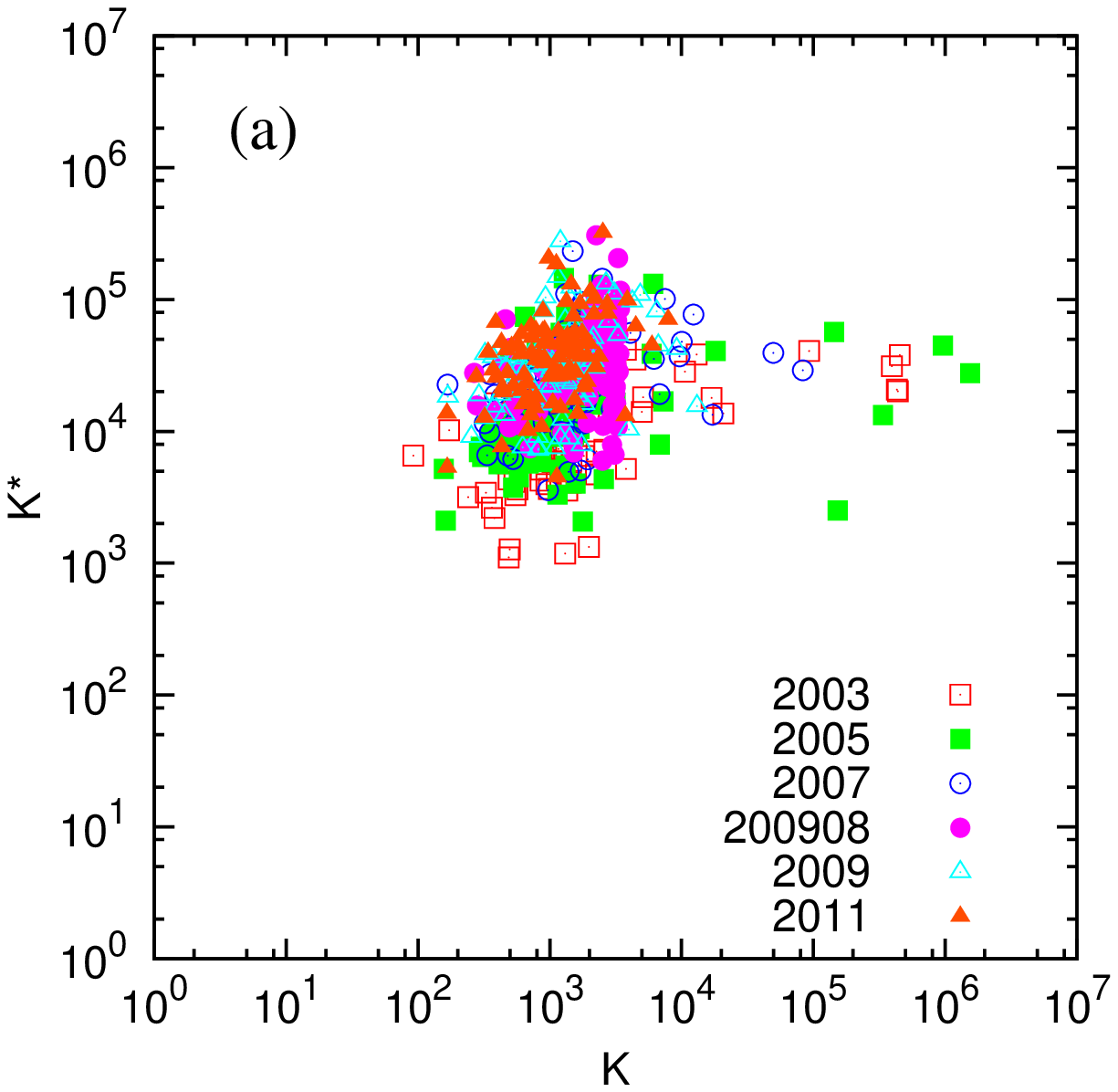}
\includegraphics[clip=true,width=3.9cm,angle=-90]{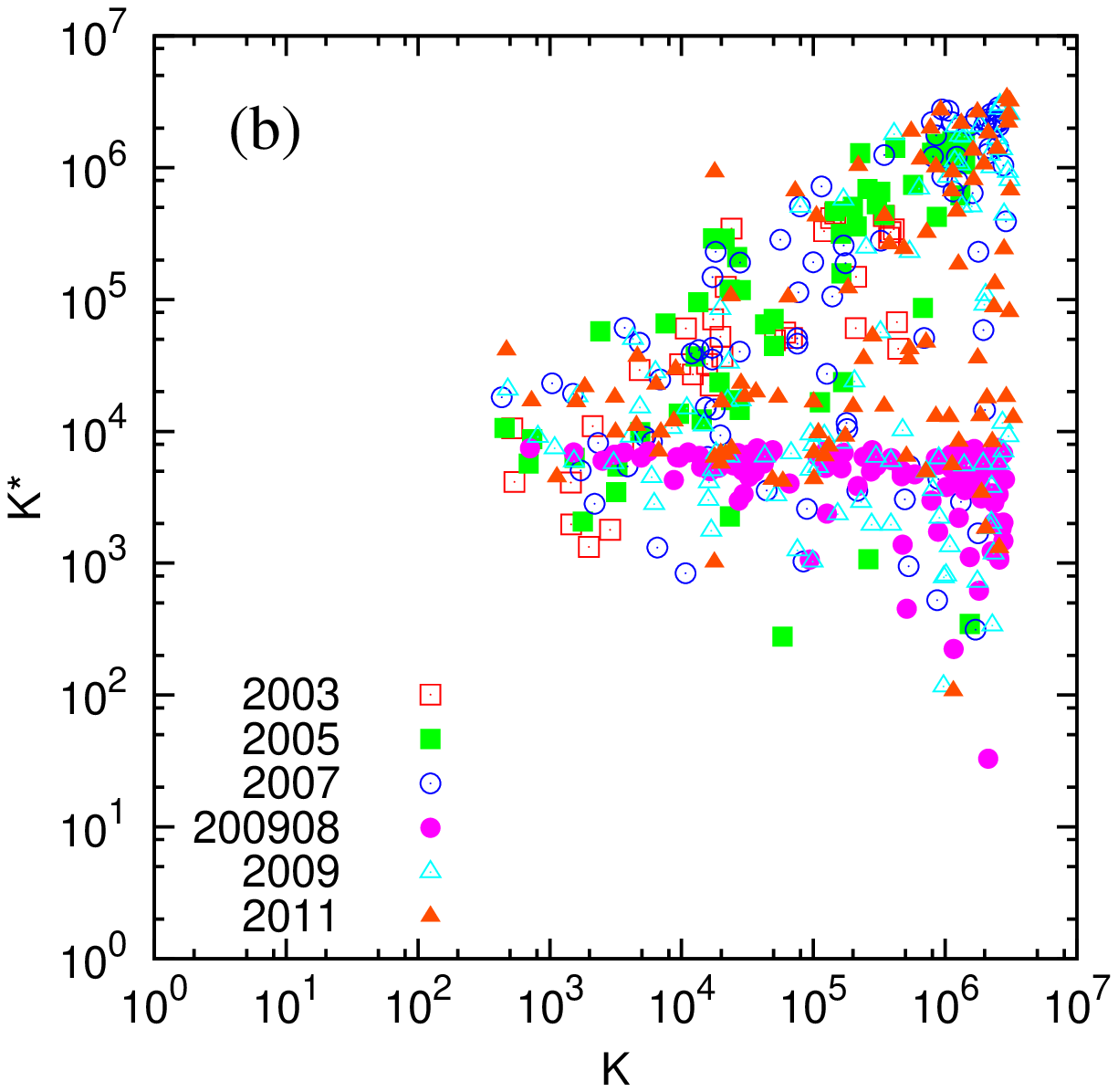}\\
\includegraphics[clip=true,width=3.9cm,angle=-90]{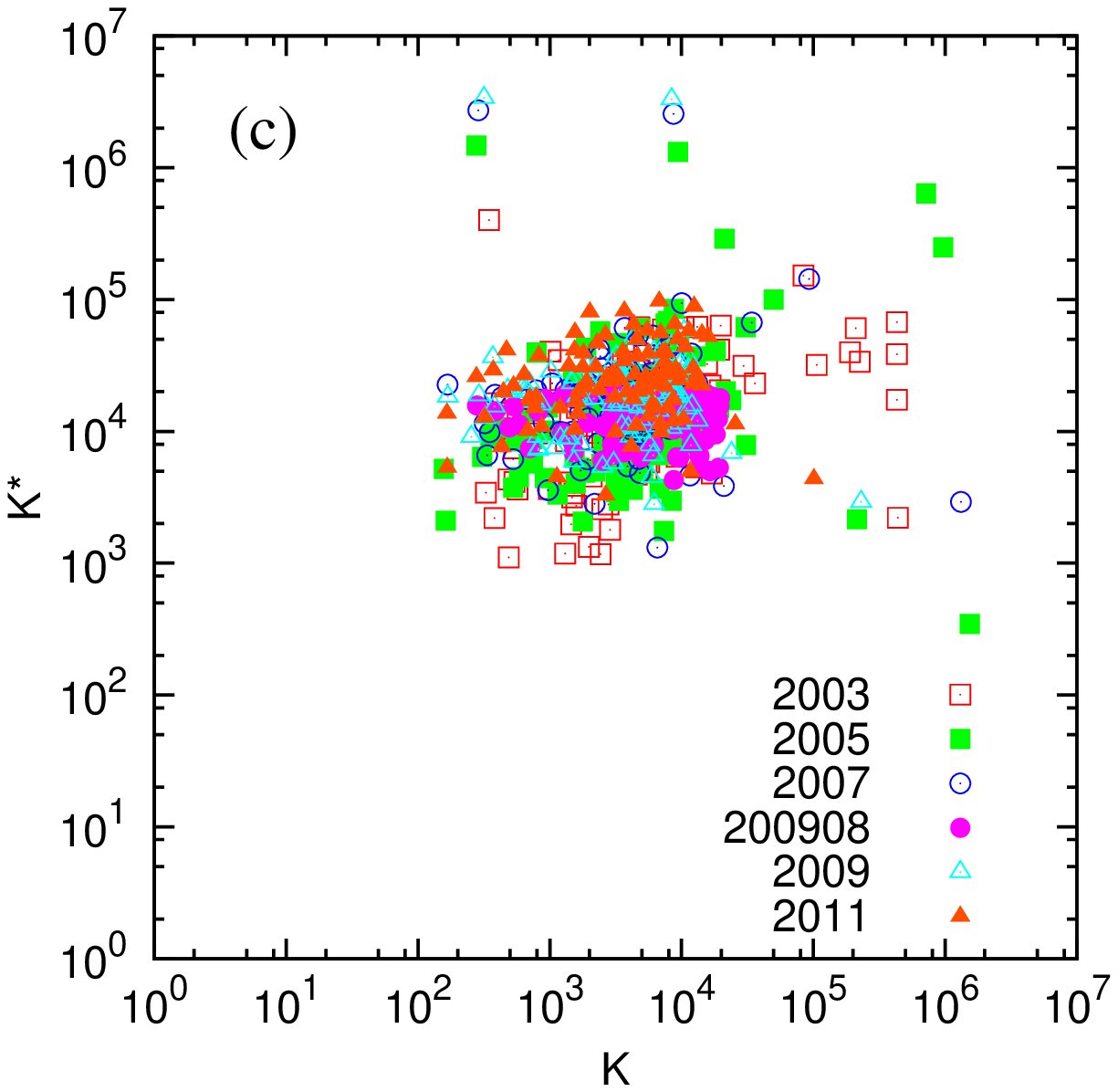}
\includegraphics[clip=true,width=3.9cm,angle=-90]{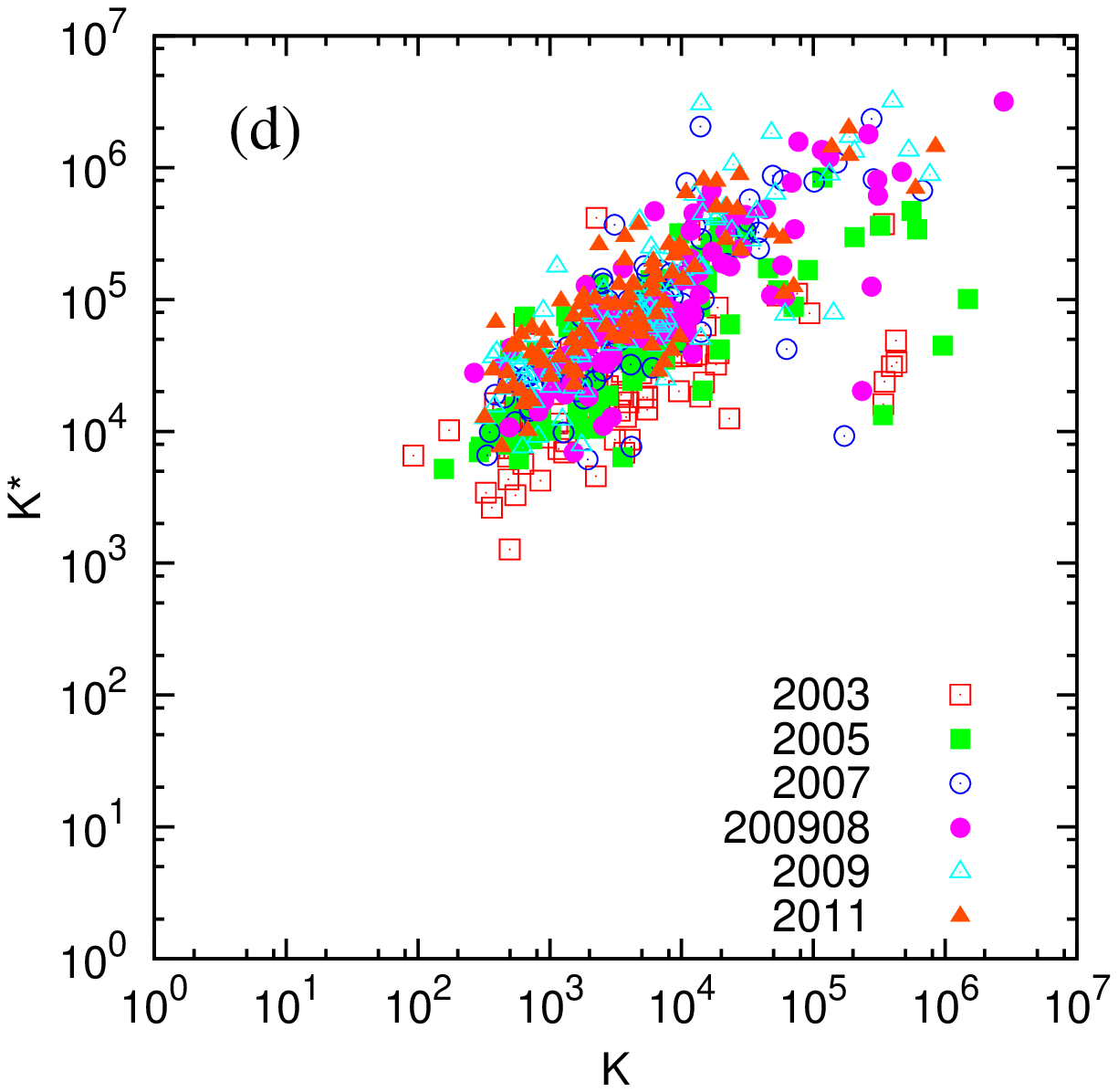}
\vglue -0.1cm
\caption{Change of locations of top-rank persons of Wikipedia in
K-K* plane. Each list of top ranks is determined by data of 
top 100 personalities of time slot 200908 in corresponding rank.
Data sets are shown for  (a) PageRank,
 (b) CheiRank, (c) 2DRank, (d) rank from Hart \cite{hart}.
}
\label{fig4}
\end{center}
\end{figure}

The distributions of personalities in 
PageRank\--CheiRank plane is shown at 
various time slots in Fig.~\ref{fig4}. 
There are visible fluctuations of distribution of nodes
for years 2003 and  2005 when the Wikipedia size
has rapid growth (see e.g. Fig.~\ref{fig4}a,c). 
For other years 2007-2011
the distribution of 
top 100 nodes of personalities of PageRank 
and 2DRank is more compact even if
individual nodes change their rank
positions in $(K,K^*)$ plane: in these years
the points form one 
compact cloud (see panels $(a,c)$ in Fig.~\ref{fig4}).
For top 100 personalities 
of CheiRank the fluctuations remain strong 
during all years (panel $(b)$ in Fig.~\ref{fig4}). 
Indeed, the number of outgoing 
links is more easy to be modified by
authors writing a given article,
while a modification of ingoing links depends on 
authors of other articles.

In Fig.~\ref{fig4}, we also show the distribution of 
top 100 personalities from Hart's book \cite{hart}  
(the list of names is also available at the web page \cite{zzswiki}).
This distribution also remains stable in years 2007-2011.
It is interesting to note that while top PageRank and 2DRank nodes 
form a kind of droplet in $(K,K^*)$ plane, the distribution
of Hart's personalities approximately follows the ridge
along the line $\ln K^* \approx \ln K + 4.6$.

\begin{figure}
\begin{center}
\includegraphics[clip=true,width=6.0cm,angle=-90]{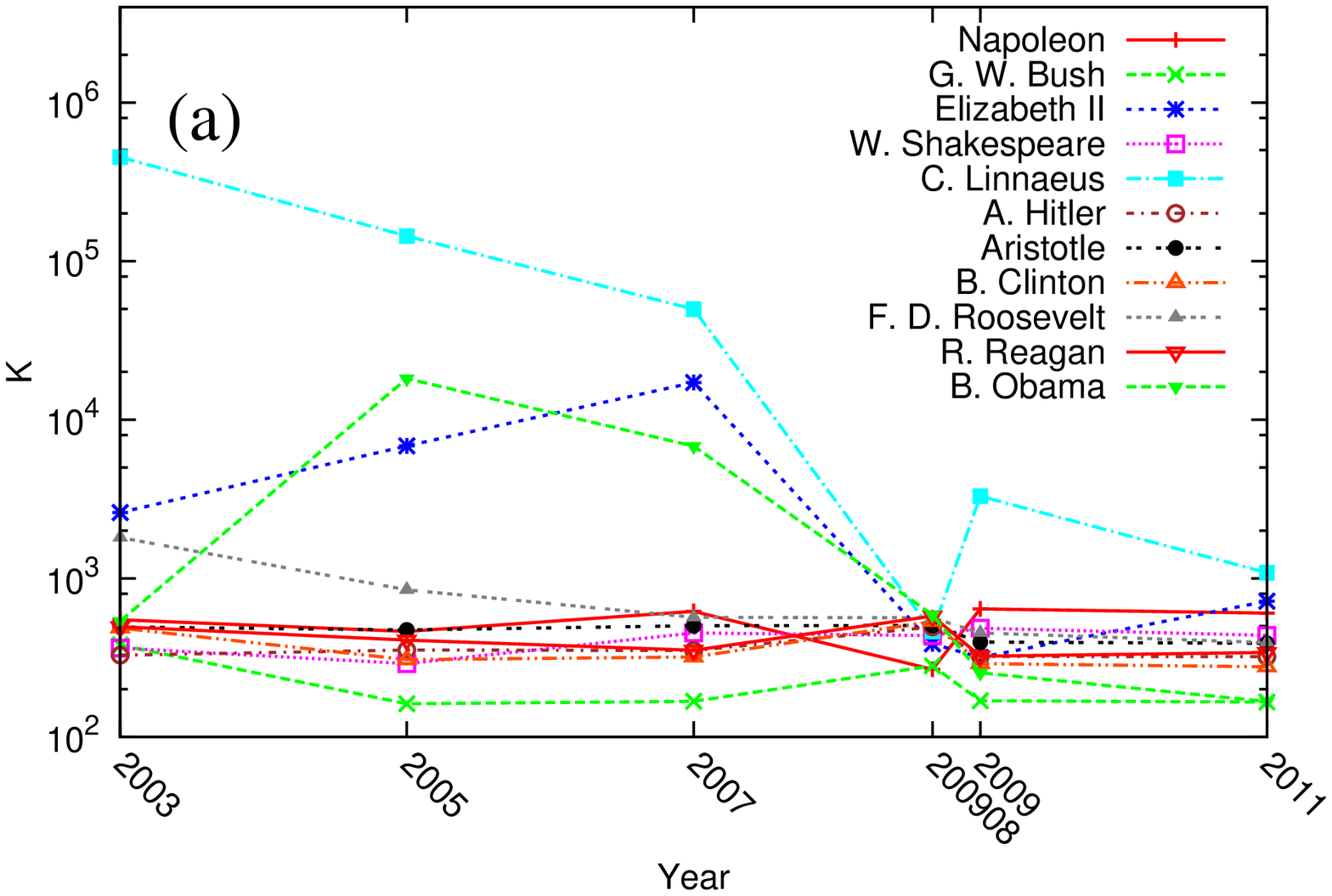}\\
\includegraphics[clip=true,width=6.0cm,angle=-90]{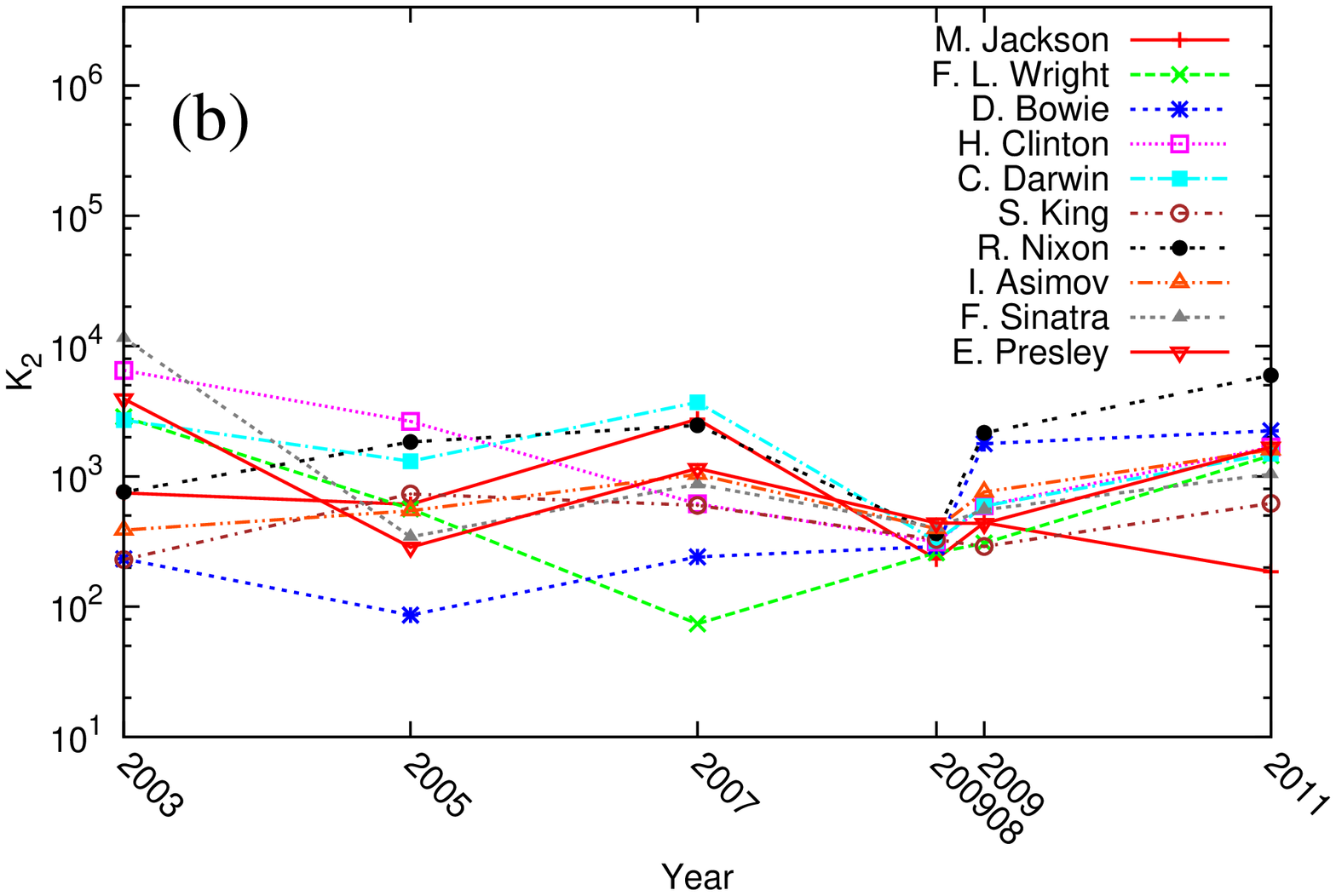}
\vglue -0.1cm
\caption{ Time evolution  of top 10 personalities of year 200908
in indexes of PageRank $K$ (a) and 2DRank $K_2$ (b); B.Obama is added in
panel (a).
}
\label{fig5}
\end{center}
\end{figure}

The time evolution of top 10 personalities of slot 200908
is shown in Fig.~\ref{fig5} for PageRank and 2DRank.
For PageRank the main part of personalities keeps
their rank position in time, e.g. G.W.Bush remains at 
first-second position. B.Obama significantly improves his ranking 
as a result of president elections. There are strong variations
for Elizabeth II which we relate to modification of
article name during the considered time interval.
We also see a steady improvement of ranking of
C.Linnaeus that we attribute to a growth of descriptions of
various botanic, insect and animal species
which quote C.Linnaeus. 
For 2DRank we observe stronger variations of $K_2$ index with time.
Such a politician as R.Nixon has increasing $K_2$ index with time
since the period of his presidency is finished more and more years ago
and events linked to his political activity, e.g. like Watergate scandal,
have lower and lower echo with time.
At the same time such representatives of arts as 
M.Jackson, F.Sinatra, and S.King remain at approximately
constant level of $K_2$ or even improve their ranking.

We note that in Fig.~\ref{fig5} the dispersion of points
increases in both directions of time from the slot 200908.
This happens because top 10 persons are taken at this moment of time
and thus as for any diffusion process the dispersion grows 
forward and backward in time. Thus we checked that if we take top 10
persons in December 2009 then the dispersion increases in both directions
of time from this point.

\begin{figure}
\begin{center}
\includegraphics[clip=true,width=3.5cm,angle=-90]{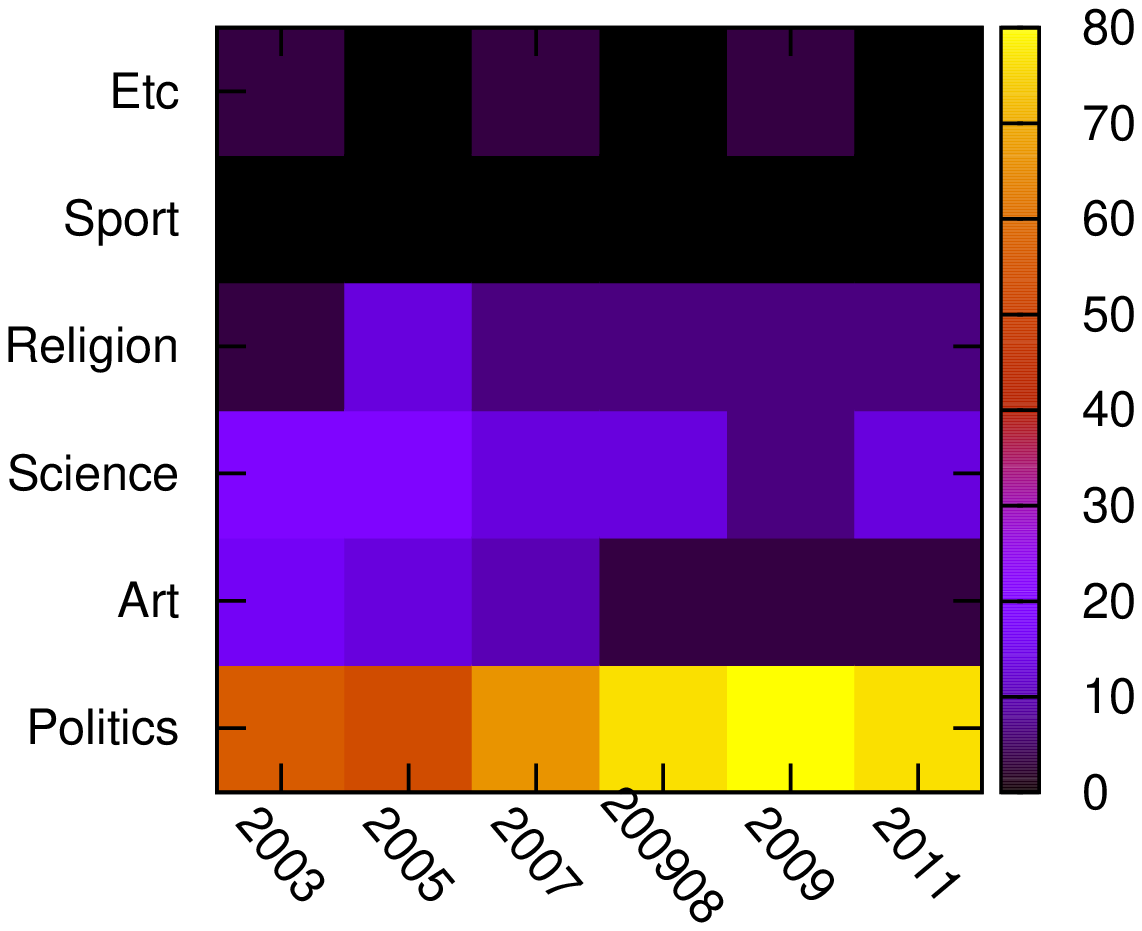}
\includegraphics[clip=true,width=3.5cm,angle=-90]{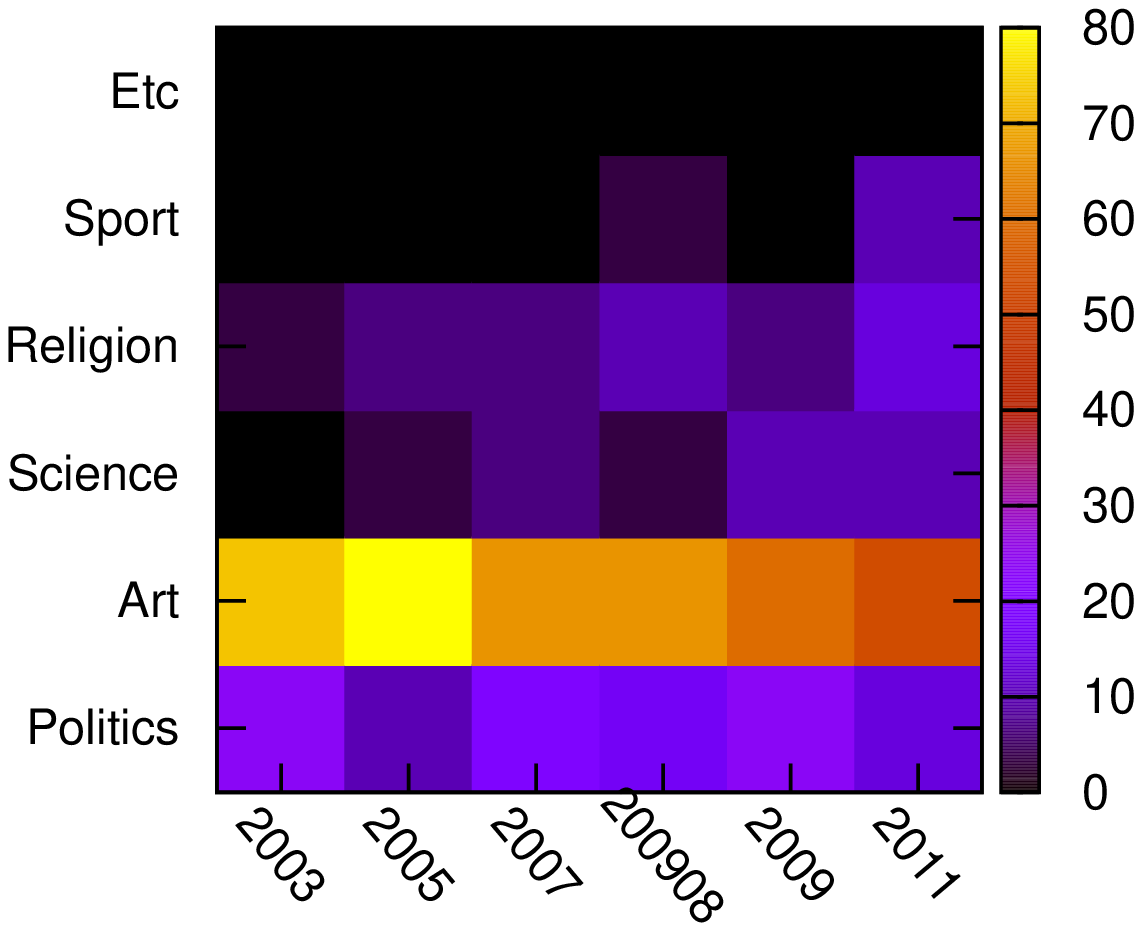}
\vglue -0.1cm
\caption{Left panel: distribution of top 30 PageRank personalities 
over 6 activity categories at various years of Wikipedia.
Right panel: distribution of top 30 2DRank personalities
over the same activity categories at same years.
Categories are  politics, art, science,  religion, sport, etc (other).
Color shows the number of personalities 
for each activity expressed in percents.
}
\label{fig6}
\end{center}
\end{figure}

In \cite{zzswiki} it was pointed out that 
the top personalities of PageRank are dominated by politicians
while for 2DRank the dominant component of human activity
is represented by artists. We analyze the time evolution
of the distribution of top 30 personalities over
6 categories of human activity 
({\it politics, arts, science, religion, sport and etc (or others)}).
We attribute a personality to an activity 
following the description of
Wikipedia article about this person
(presidents, kings, imperators belong to politics,
artists, singers, composers, painters belong to arts
scientists and philosophers to science,
priests and popes to religion, sportsmen to sport,
etc includes all other activities not listed above).
In fact, the category {\it etc} contains only C.Columbus. 
The results are presented in Fig.~\ref{fig6}.
They clearly show that 
the PageRank personalities are dominated by politicians
whose percentage increases with time, while 
the percent of arts decreases. For 2DRank we see that
the arts are dominant even if their percentage decreases with time.
We also see the  appearance of sport which is absent in PageRank.
The mechanism of the qualitative ranking differences between
two ranks is related to the fact that 2DRank 
takes into account via CheiRank a contribution of outgoing links.
Due to that singers, actors, sportsmen improve their 
CheiRank and 2DRrank  positions since articles about them 
contain various music albums,
movies and sport competitions with many outgoing links. 
Due to that the component of
arts gets higher positions in 2DRank in contrast to politics dominance
in PageRank. Thus the two-dimensional ranking on PageRank-CheiRank plane
allows to select qualities of nodes according to 
their popularity and communicativity. 

\section{Ranking of universities}

\begin{figure}
\begin{center}
\includegraphics[clip=true,width=7.6cm,angle=-90]{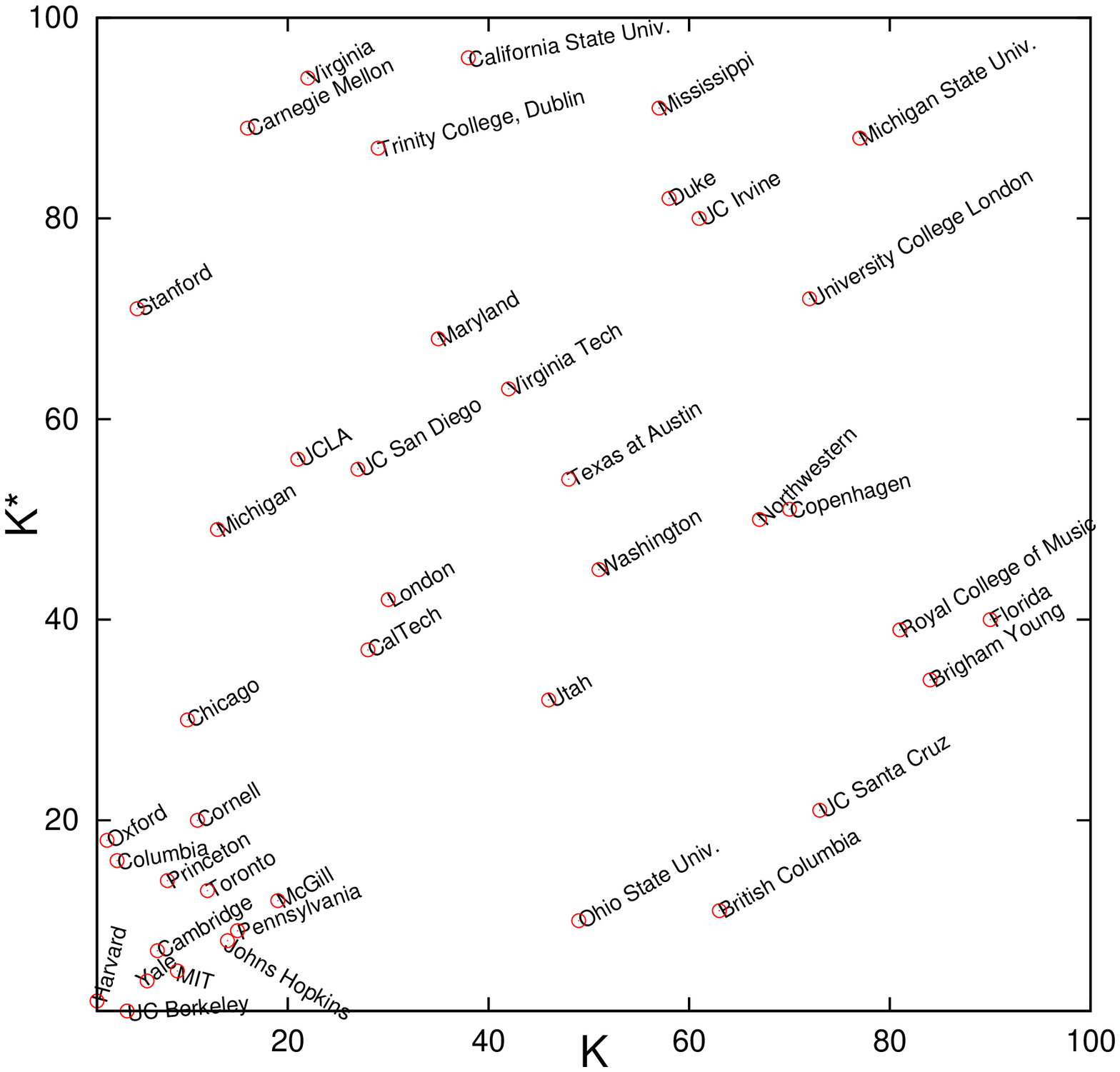}\\
\includegraphics[clip=true,width=7.6cm,angle=-90]{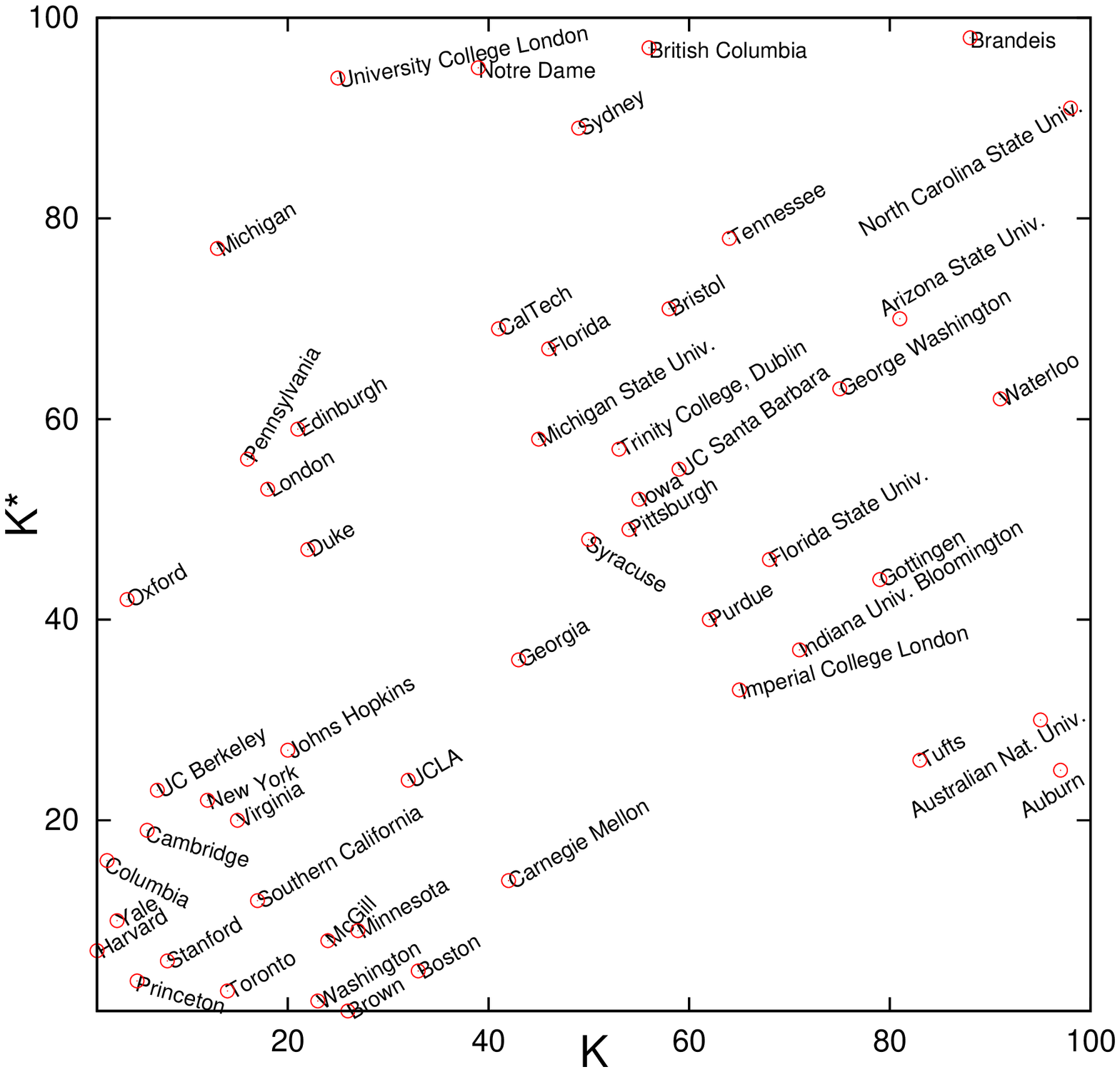}\\
\includegraphics[clip=true,width=7.6cm,angle=-90]{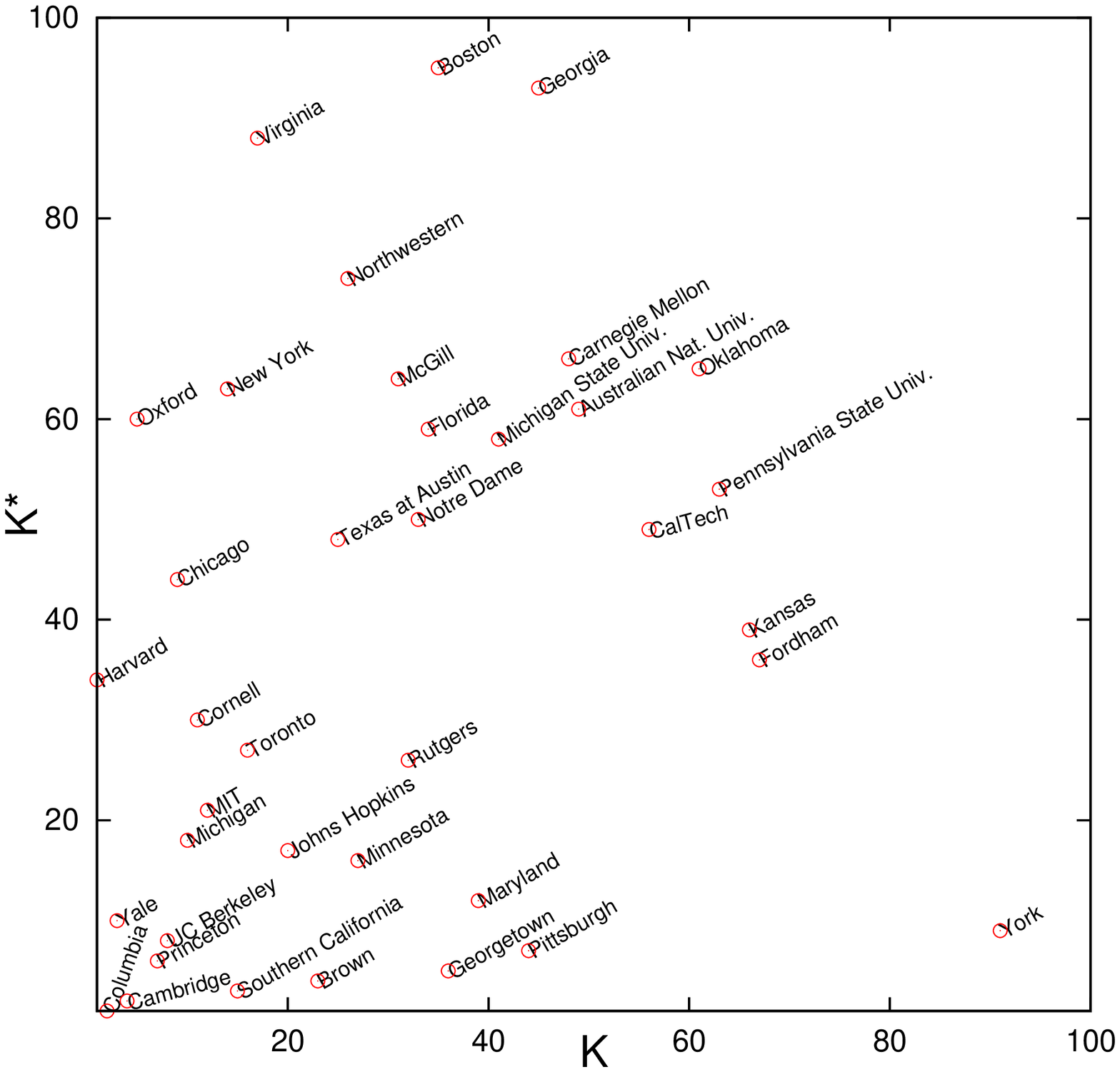}
\vglue -0.1cm
\caption{University of Wikipedia articles 
in the local CheiRank versus PageRank plane at different years;
panels are for years 2003, 2005, 2007
(from top to bottom).
}
\label{fig7}
\end{center}
\end{figure}

\begin{figure}
\begin{center}
\includegraphics[clip=true,width=7.6cm,angle=-90]{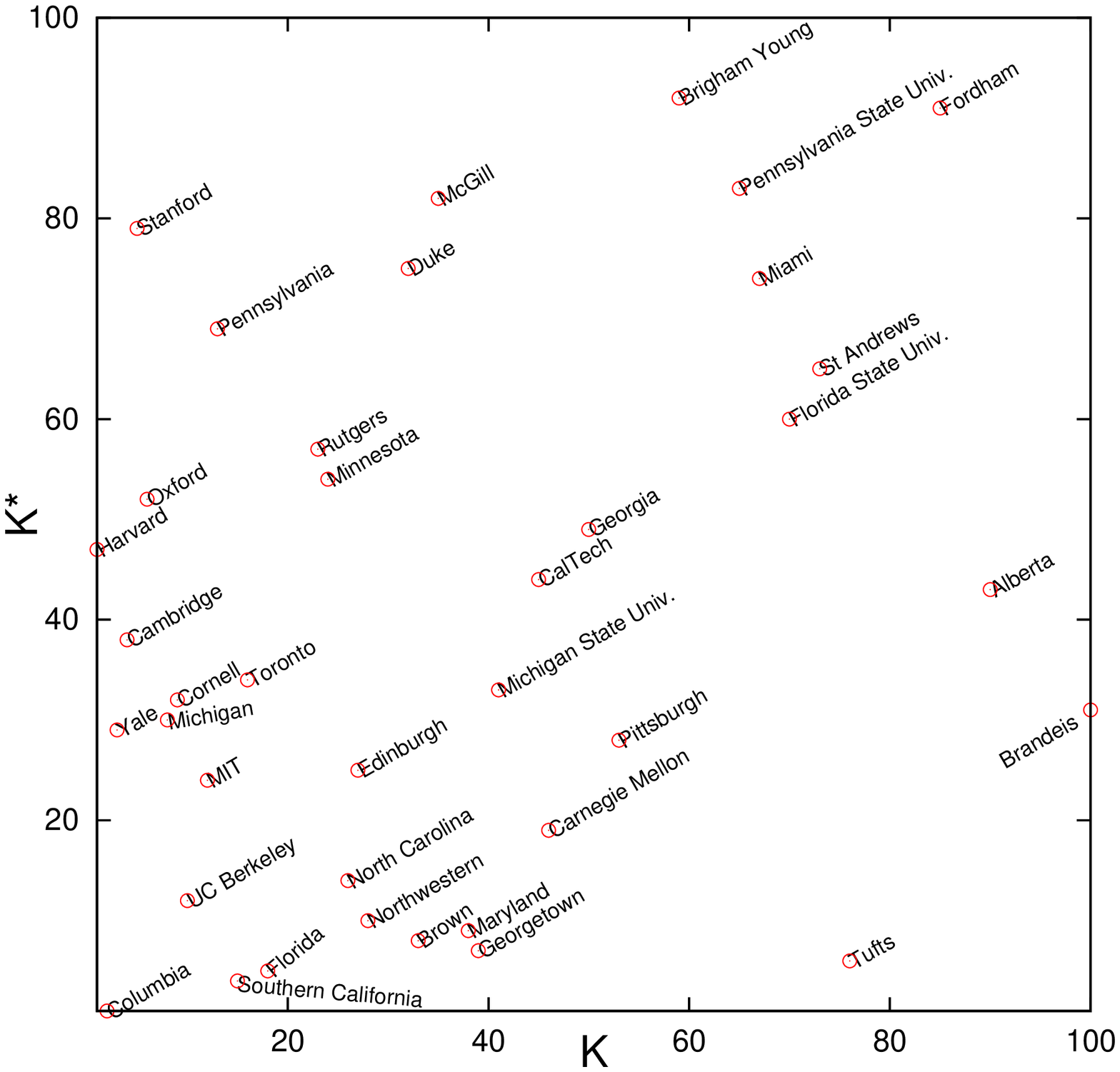}\\
\includegraphics[clip=true,width=7.6cm,angle=-90]{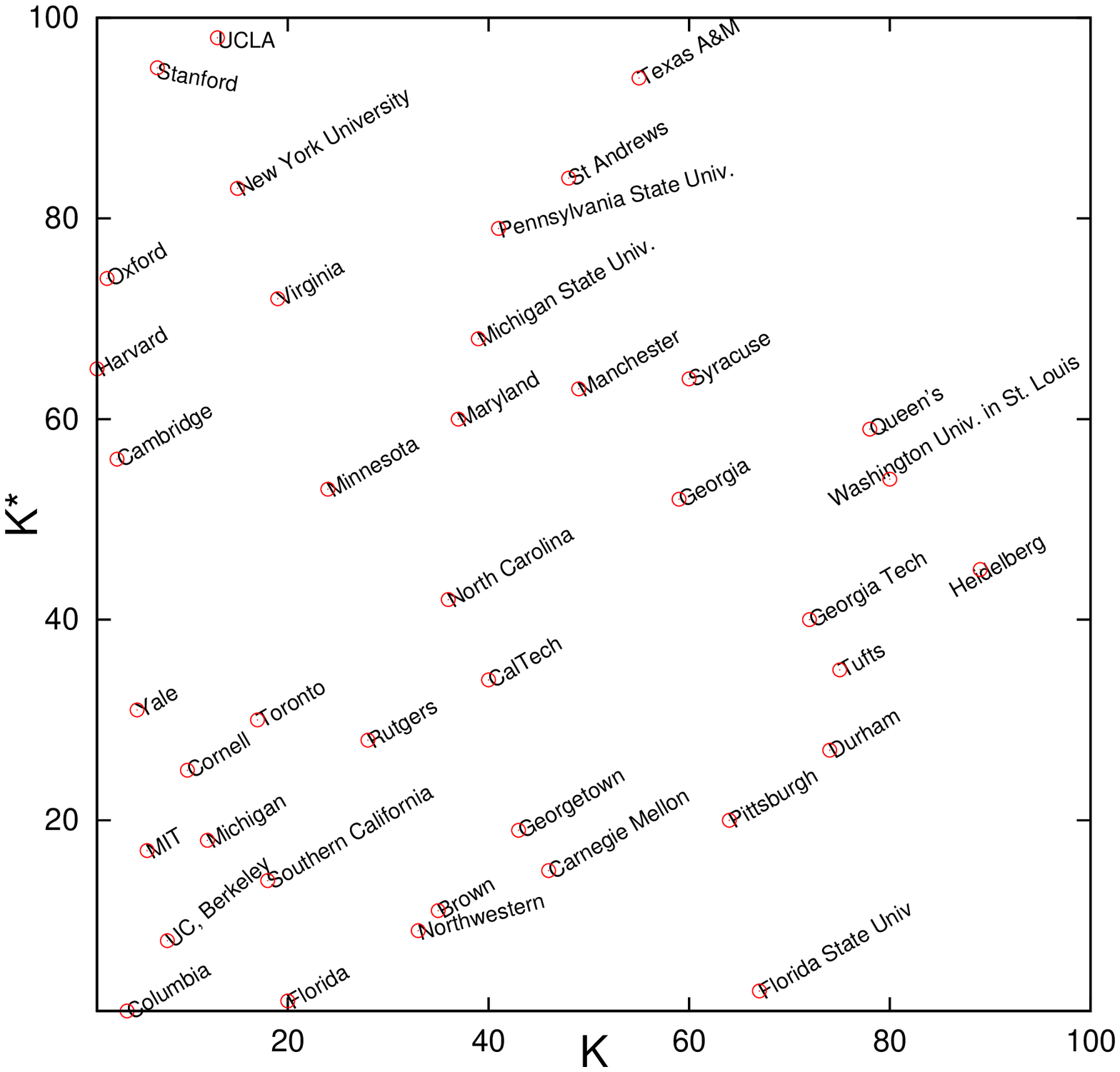}\\
\includegraphics[clip=true,width=7.6cm,angle=-90]{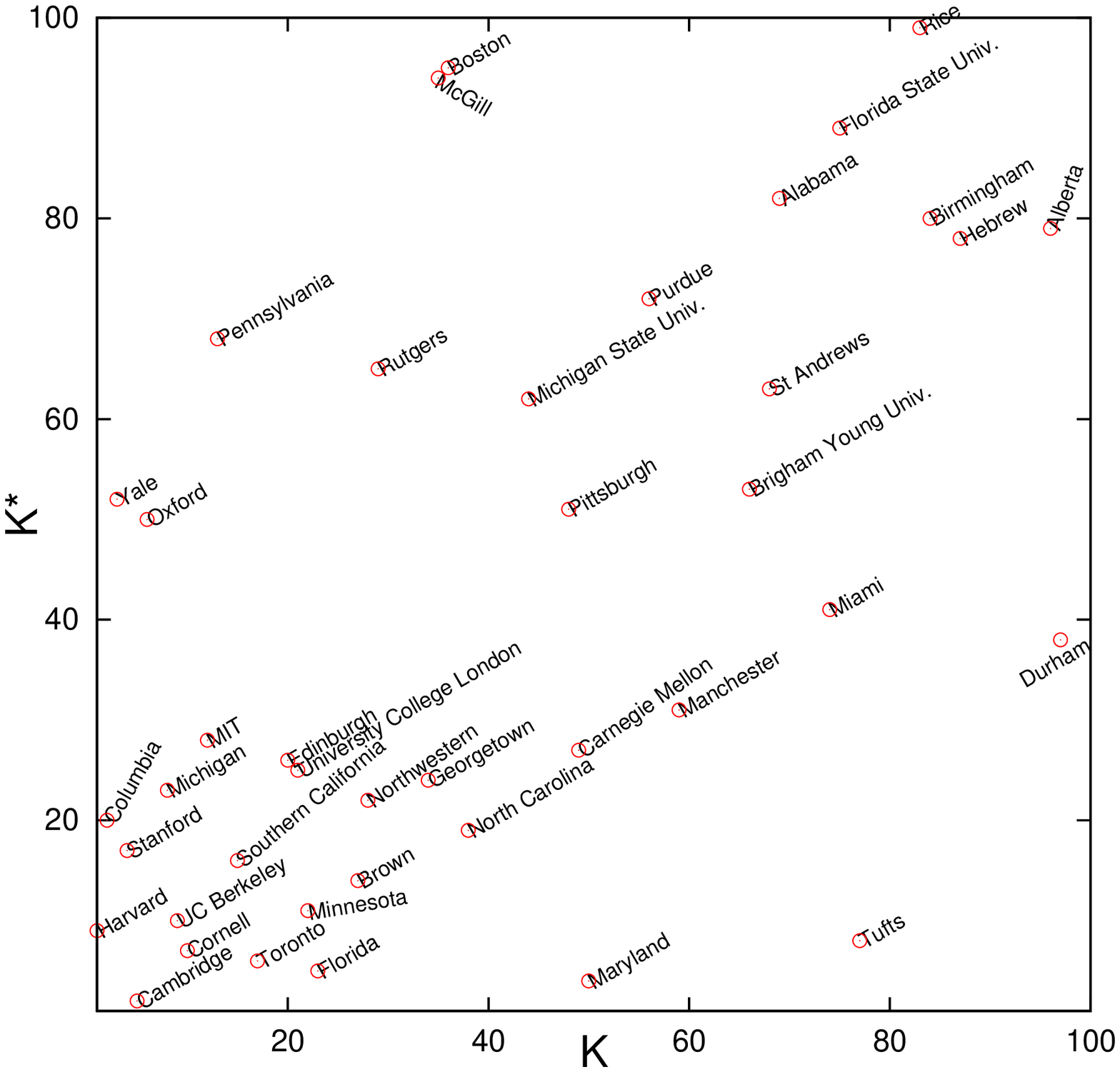}
\vglue -0.1cm
\caption{Same as in Fig.~\ref{fig7} 
for years 2009, 200908, 2011
(from top to bottom).
}
\label{fig8}
\end{center}
\end{figure}

The local ranking of top 100 universities is shown in Fig.~\ref{fig7}
for years 2003, 2005, 2007 and in  Fig.~\ref{fig8}
for 2009, 200908, 2011. The local ranking is obtained
by selecting top 100 universities appearing in PageRank list
so that they get their university ranking $K$ from 1 to 100.
The same procedure is done for CheiRank list of universities
obtaining their local CheiRank index $K^*$ from 1 to 100.
Those universities which enter inside $100 \times 100$ 
square on the local index plane $(K,K^*)$ are shown in
Figs.~\ref{fig7},~\ref{fig8}.

\begin{figure}
\begin{center}
\includegraphics[clip=true,width=6.0cm,angle=-90]{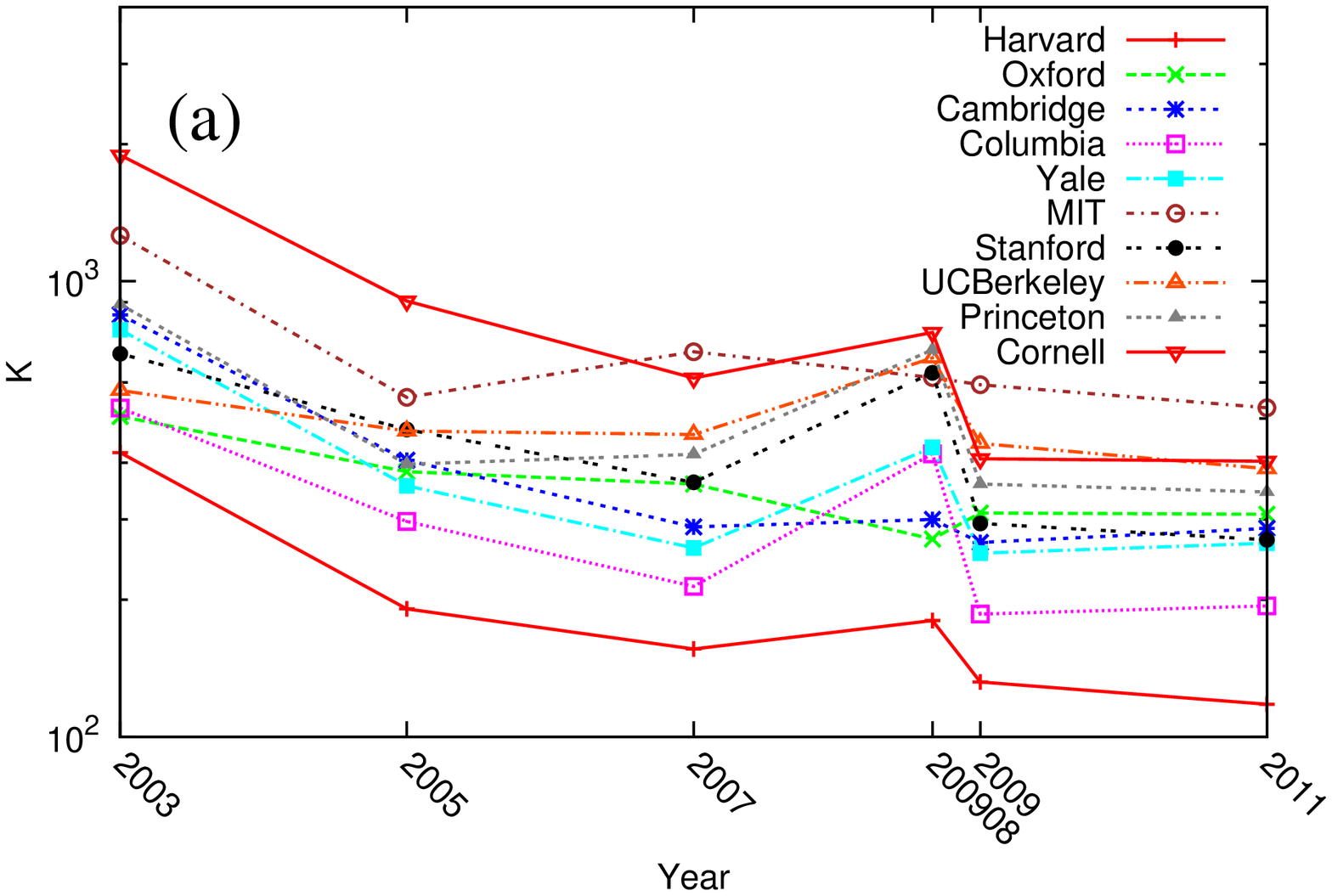}\\
\includegraphics[clip=true,width=6.0cm,angle=-90]{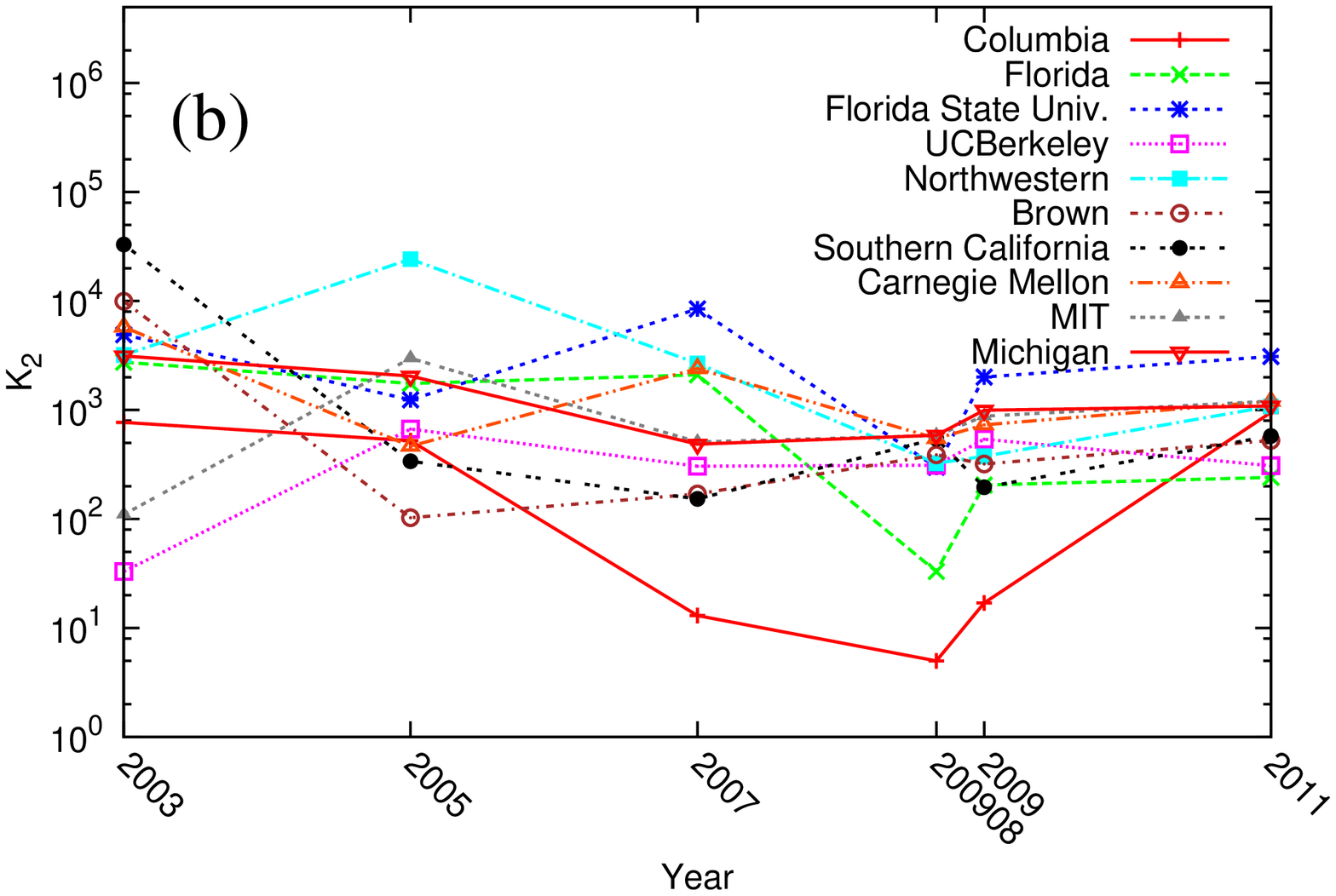}
\vglue -0.1cm
\caption{Time evolution of global ranking of top 10 Universities
of year 200908 in indexes of
PageRank $K$ (a)
and  2DRank $K_2$ (b).
}
\label{fig9}
\end{center}
\end{figure}

The data show that the top PageRank universities are
rather stable in time, e.g. Harvard is always on the first top
position, Columbia at the second position
and Yale is the third for the majority of  years. 
Also there is a relatively small number of intersection of 
curves of $K$ with years.
At the same time the positions in $K_2$ and $K^*$
are strongly changing in time. To understand the origin of this
variations in CheiRank we consider the case of U Cambridge.
Its Wikipedia article in 2003 is rather short
but it contains the list of all 31 Colleges with direct links to their
corresponding articles. This leads to a high position of U Cambridge
with university $K^*=4$ in 2003 (Fig.~\ref{fig9}).
However, with time the direct links remain only to 
about 10 Colleges  while the whole number of Colleges 
are presented by a list of names without links.
This leads to a significant increase of index up to $K^* \approx 40$ 
at Dec 2009. However, at Dec 2011 U Cambridge
again improves significantly its CheiRank obtaining $K^*=2$. 
The main reason of that is the 
appearance of section of ``Notable alumni and academics'' which 
provides direct links to articles about outstanding scientists 
studied and/or worked at U Cambridge
that leads to second position at $K^*=2$
among all universities. We note that in 2011
the top CheiRank University is
George Mason University with university $K^*=1$.
The main reason of this high ranking is the presence of detailed lists
of alumni in politics, media, sport
with direct links to articles about corresponding personalities
(including former director of CIA). These two examples show that
the links, kept by a university with a large number of
its alumni, significantly increase CheiRank
position of university. We note that colleges specialized in arts,
religion, politics usually preserve more links with their alumni
as also was pointed in \cite{zzswiki}.

The time evolution of global ranking of top 10 universities
of year 200908 for PageRank and 2DRank is shown in Fig.~\ref{fig9}.
The results show the stability of PageRank order
with a clear tendency of top universities (e.g. Harvard)
to go with time to higher and higher top positions of $K$.
Thus for Harvard the global value of $K$ changes from $K \approx 300$
in 2003 to $K \approx 100$ in 2011, 
while the whole size $N$ of the Wikipedia 
network increases almost by a factor $10$ during this time interval.  
Since Wikipedia ranks all human knowledge, the stable improvement
of PageRank indexes of universities reflects the global
growing importance of universities in the world of human activity
and knowledge.

The time evolution of  top 10 universities of year 200908 in 2DRank
remains on average approximately at a constant level $K_2 \approx const$ 
in time (without the above global improvement visible for
for PageRank), it also  shows more interchanges of ranking order
comparing to PageRank case.
We think that an example of U Cambridge considered above
explains the main reasons of these fluctuations. 
In view of 10 times increase of the whole network size 
during the period 2003 - 2011
the average stability of 2DRank of universities
also confirms the 
significant importance of their place in human activity. 

Finally we compare the Wikipedia ranking of universities 
in their local PageRank index $K$ with those of Shanghai university ranking
\cite{shanghai}. In the top 10 of Shanghai university rank
the Wikipedia PageRank recovers 9 (2003), 9 (2005),
8 (2007), 7 (2009), 7 (2011). Thus on average
the Wikipedia PageRanking of universities recovers $80\%$
of top universities of Shanghai ranking during the considered time period.
This shows that the
Wikipedia ranking of universities gives the results being
rather similar to Shanghai ranking performed 
on the basis of other selection criteria.
A small decrease of overlap with time
can be attributed to earlier launched activity of
leading universities on Wikipedia.

\section{Google matrix spectrum}

Finally we discuss the time evolution of the spectrum
of Wikipedia Google matrix taken at $\alpha=1$.
We perform the numerical diagonalization based on the Arnoldi method
\cite{arnoldibook,golub} using the additional improvements
described in \cite{ulamfrahm,univuk} with the  Arnold dimension $n_A=6000$.
The Google matrix is reduced to the form
\begin{equation}
\label{eq4}
S=\left(\begin{array}{cc}
S_{ss} & S_{sc}  \\
0 & S_{cc}\\
\end{array}\right)\;
\end{equation}
where $S_{ss}$ describes disjoint subspaces $V_j$ of dimension 
$d_j$ invariant by applications of $S$;  $S_{cc}$
depicts the remaining part of nodes 
forming the wholly connected {\it core space}.
We note that $S_{ss}$ is by itself composed of many small diagonal blocks for 
each invariant subspace and hence those eigenvalues can be efficiently obtained 
by direct (``exact'') numerical diagonalization.
The total subspace size $N_s$, the number of independent subspaces $N_d$,
the maximal subspace dimension $d_{\rm max}$ and the number $N_1$
of $S$ eigenvalues with $\lambda=1$ are given in Table 2
(See also Appendix).
The spectrum and eigenstates of the core space $S_{cc}$
are determined by the Arnoldi method with Arnoldi dimension $n_A$ 
giving the eigenvalues $\lambda_i$ of $S_{cc}$ with largest modulus.
Due to the finite value of $n_A$ available for numerical simulations
eigenvalues with small $|\lambda_i|$ are not computed that leaves 
an empty space in the complex plane $\lambda$
(see discussions  in \cite{twitter,wikispectrum}). 
Here we restrict ourselves to the statistical analysis of the 
spectrum $\lambda_i$. The analysis of eigenstates $\psi_i$
($G \psi_i = \lambda_i \psi_i$),
which has been done in \cite{wikispectrum} for the slot 200908,
is left for future studies for other time slots.

\begin{figure}
\begin{center}
\includegraphics[clip=true,width=7.7cm]{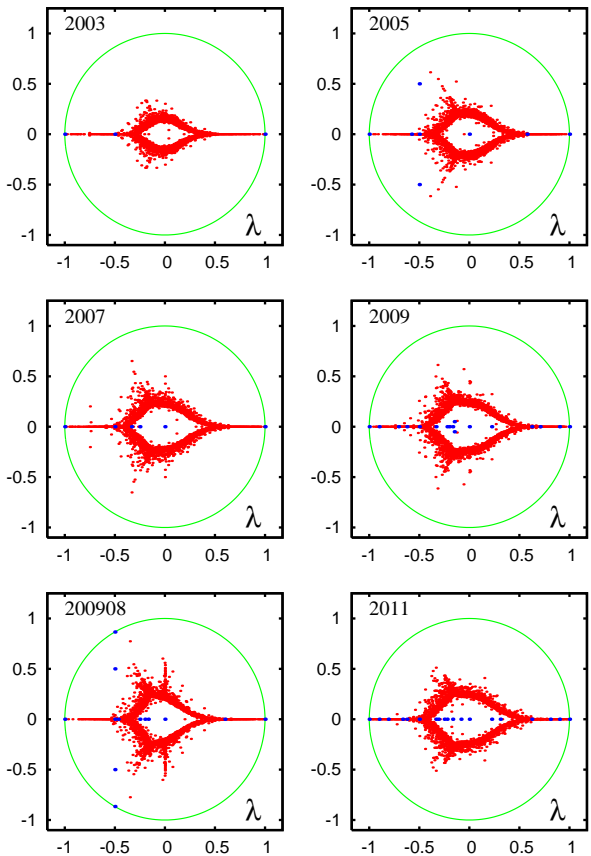}
\vglue -0.1cm
\caption{Spectrum of eigenvalues $\lambda$ of the Google matrix $G$ of 
Wikipedia at different years shown at $\alpha=1$. 
Red dots are core space eigenvalues, blue dots are subspace 
eigenvalues and the full green curve shows the unit circle. 
The core space eigenvalues were calculated 
by the projected Arnoldi method with 
Arnoldi dimensions $n_A=6000$.
}
\label{fig10}
\end{center}
\end{figure}

\begin{figure}
\begin{center}
\includegraphics[clip=true,width=7.7cm]{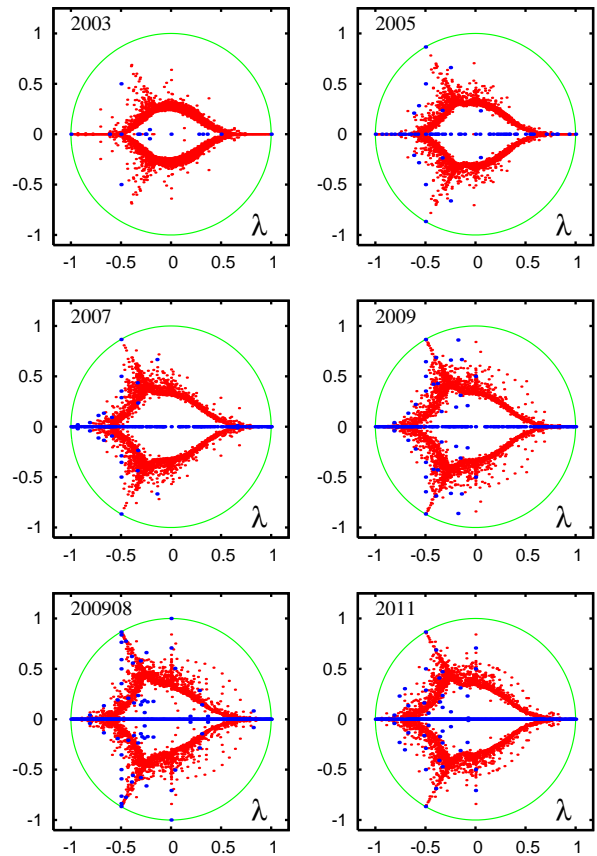}
\vglue -0.1cm
\caption{Same as in Fig.~\ref{fig10} but for the spectrum of matrix $G^*$.
}
\label{fig11}
\end{center}
\end{figure}

The spectrum for all Wikipedia time slots
is shown in Fig.~\ref{fig10} for $G$
and in Fig.~\ref{fig11} for $G^*$. We see that the spectrum remains stable
for the period 2007 - 2011 even if there is a small
difference of slot 200908 due to a slightly different 
cleaning link procedure (see Appendix).
For the spectrum of $G^*$ in 2007 - 2011 we observe a well pronounced
star structure which can be recognized as a composition
of triplet and quadruplet leaves (triangle and cross). 
This structure is very similar
to those found in random unistochastic
and orthostochastic matrices of size $N=3$ and $4$
\cite{karol2003} (see Fig.4 therein). 
This fact has been pointed in \cite{wikispectrum}
for the slot 200908. Now we see that this is a generic phenomenon
which remains stable in time. This indicates that there
are dominant  groups of 3-4 nodes which have 
structure similar to random unistochastic or orthostochastic
matrices with strong ties between 3-4 nodes and various random permutations
with random hidden complex phases. 
The spectral star structure is significantly more
pronounce for the case of $G^*$ matrix.
We attribute this to more significant fluctuations of 
outgoing links that probably makes sectors of $G^*$
to be more similar to elements of unistochastic matrices.
A further detailed analysis will be useful to understand
this star structure and its links with
various communities inside Wikipedia.

As it is shown in \cite{wikispectrum}
the eigenstates of $G$ and $G^*$ select certain well-defined
communities of the Wikipedia network. Such 
an eigenvector detection of the communities 
provides a new method of communities detection
in addition to more standard methods developed in
network science and described in \cite{fortunato}. 
However, the analysis of eigenvectors
represents a separate detailed research 
and in this work we restrict ourselves
to PageRank and CheiRank vectors.

Finally we note that the fraction of isolated subspaces
is very small for $G$ matrix. It is increased 
approximately by a factor 10 for $G^*$
but still it remains very small compared to the
networks of UK universities analyzed in \cite{univuk}.
This fact reflects  a strong connectivity of network of
Wikipedia articles. 

\section{Discussion}

In this work we analyzed the time evolution of ranking
of network of English Wikipedia articles. 
Our study demonstrates the stability of such  
statistical properties as PageRank and CheiRank probabilities,
the article density distribution in PageRank-CheiRank plane
during the period 2007 - 2011.
The analysis of human activities
in different categories shows  
that PageRank gives main accent to politics
while the combined 2DRank gives more
importance to arts. We find that with time the
number of politicians in the top positions
increases. Our analysis of ranking of universities
shows that on average the global ranking of
top universities goes to higher and higher positions.
This clearly marks the growing importance of
universities for the whole range of human activities 
and knowledge. 
We find that Wikipedia PageRank 
recovers 70 - 80 \% of top 10 universities
from Shanghai  ranking \cite{shanghai}.
This confirms the reliability of Wikipedia ranking.  

We also find that the spectral structure of the Wikipedia
Google matrix remains stable during the time period
2007 -2011 and show that its arrow star structure 
reflects certain features of small size unistochastic matrices.
 
{\bf Acknowledgments:}
Our research presented here is supported 
in part by the EC FET Open project 
``New tools and algorithms for directed network analysis''
(NADINE $No$ 288956). This work was granted access 
to the HPC resources of 
CALMIP (Toulouse) under the allocations 2012-P0110, 2013-P0110.
We also acknowledge the France-Armenia collaboration grant 
CNRS/SCS $No$ 24943 (IE-017) on ``Classical and quantum chaos''.

\renewcommand{\theequation}{A-\arabic{equation}}
  \setcounter{equation}{0}  
\renewcommand{\thefigure}{A-\arabic{figure}}
  \setcounter{figure}{0}  
\section{Appendix}

The tables with all network parameters used in this work
are given tables 1 and 2.
The notations used in the tables are:
$N$ is network size, $N_\ell$ is the number of links,
$n_A$ is the Arnoldi dimension used for the Arnoldi method for the 
core space eigenvalues, $N_d$ is the number of invariant subspaces,
$d_{\rm max}$ gives a maximal subspace dimension,
$N_{\rm circ.}$ notes number of eigenvalues on the unit circle with 
$|\lambda_i|=1$, $N_1$ notes 
number of unit eigenvalues with $\lambda_i=1$.
We remark that $N_s\ge N_{\rm circ.}\ge N_1\ge N_d$ and $N_s\ge d_{\rm max}$.
The data for $G$ are marked by the corresponding year
of the time slot, the data for $G^*$ are marked by
the year with a star. Links cleaning procedure eliminates
all redirects for sets of 2003, 2005, 2007, 2009, 2011
(nodes with one outgoing link are eliminated; 
thus practically all redirects are eliminated, 
we do no relink articles via redirects.).
Also all articles which titlws have  only numbers or/and 
only special symbols
have been eliminated.
The set 200908 is taken from \cite{zzswiki}
where the cleaning procedure was slightly different:
all  nodes with one outgoing link were eliminated but 
no special cleaning on article titles with numbers was affected. 
This is probably the reason why 
$N_l$ of 200908 is larger than its values in
Dec 2009 and 2011.
All data sets and high resolution figures are available at
the web page \cite{webpage}.


\medskip
\begin{table}\label{table1}
\centering
\begin{tabular}{|l|c|c|c|}
\hline
& $N$ & $N_\ell$ & $n_A$ \\
\hline
\hline
\ 2003\ & \ 455436\ & \ 2033173\ & \ 6000\ \\
\hline
\ 2005\ & \ 1635882\ & \ 11569195\ & \ 6000\ \\
\hline
\ 2007\ & \ 2902764\ & \ 34776800\ & \ 6000\ \\
\hline
\ 2009\ & \ 3484341\ & \ 52846242\ & \ 6000\ \\
\hline
\ 200908\ & \ 3282257\ & \ 71012307\ & \ 6000\ \\
\hline
\ 2011\ & \ 3721339\ & \ 66454329\ & \ 6000\ \\
\hline
\end{tabular}
\caption{Parameters of all Wikipedia networks at different years
considered in the paper; set 2009 corresponds to Dec 2009, set 200908 
to Aug 2009.}
\end{table}

\medskip
\begin{table}\label{table2}
\centering
\begin{tabular}{|l|c|c|c|c|c|}
\hline
& $N_s$ & $N_d$ & $d_{\rm max}$ & $N_{\rm circ.}$ & $N_1$\\
\hline
\hline
\ 2003\ & \ 15\ & \ 7\ & \ 3\ &\ 11\ & \ 7\ \\
\hline
\ 2003$^*$\ & \ 940\ & \ 162\ & \ 60\ &\ 265\ & \ 163\ \\
\hline
\ 2005\ & \ 152\ & \ 97\ & \ 4\ &\ 121\ & \ 97\ \\
\hline
\ 2005$^*$\ & \ 5966\ & \ 1455\ & \ 1997\ &\ 2205\ & \ 1458\ \\
\hline
\ 2007\ & \ 261\ & \ 150\ & \ 6\ &\ 209\ & \ 150\ \\
\hline
\ 2007$^*$\ & \ 10234\ & \ 3557\ & \ 605\ &\ 5858\ & \ 3569\ \\
\hline
\ 2009\ & \ 285\ & \ 121\ & \ 8\ &\ 205\ & \ 121\ \\
\hline
\ 2009$^*$\ & \ 11423\ & \ 4205\ & \ 134\ &\ 7646\ & \ 4221\ \\
\hline
\ 200908\ & \ 515\ & \ 255\ & \ 11\ &\ 381\ & \ 255\ \\
\hline
\ 200908$^*$\ & \ 21198\ & \ 5355\ & \ 717\ &\ 8968\ & \ 5365\ \\
\hline
\ 2011\ & \ 323\ & \ 131\ & \ 8\ &\ 222\ & \ 131\ \\
\hline
\ 2011$^*$\ & \ 14500\ & \ 4637\ & \ 1323\ &\ 8591\ & \ 4673\ \\
\hline
\end{tabular}
\caption{$G$ and $G^*$ eigespectrum parameters for all Wikipedia networks,
year marks spectrum of $G$, year with star marks spectrum of $G^*$.}
\end{table}


\end{document}